\documentclass[aps,nopacs,superscriptaddress,12pt,tightenlines]{revtex4}

\usepackage[english]{babel}
\usepackage[pdftex]{graphicx}
\usepackage{bm}
\usepackage{hyperref}
\usepackage{amsmath}
\usepackage{amssymb}
\usepackage{bm}
\usepackage[usenames]{color}
\usepackage{wrapfig}
\usepackage{empheq}
\usepackage{scrtime}
\usepackage{setspace}

\advance\voffset -.25in

\newcommand*{\defeq}{\mathrel{\vcenter{\baselineskip0.5ex \lineskiplimit0pt
                     \hbox{\scriptsize.}\hbox{\scriptsize.}}}%
                     =}
                     
\begin{document}

\title{A high-performance analog Max-SAT solver and its application to Ramsey numbers} 

\author{Botond Moln\'ar}
\affiliation{Faculty of Physics, Babe\c{s}-Bolyai University, Cluj-Napoca, 400084  Romania}
\affiliation{Romanian Institute of Science and Technology, Cluj-Napoca, 400487 Romania}
\author{Melinda Varga}
\affiliation{Department of Physics and the Interdisciplinary Center for Network Science and Applications, University of Notre Dame, Notre Dame, IN, 46556 USA}
\affiliation{Biochemistry and Molecular Biology Program, The College of Wooster, Wooster, OH, 44691 USA}
\author{Zolt\'an Toroczkai}\email{toro@nd.edu}
\affiliation{Department of Physics and the Interdisciplinary Center for Network Science and Applications, University of Notre Dame, Notre Dame, IN, 46556 USA}
\author{M\'aria Ercsey-Ravasz}\email{ercsey.ravasz@phys.ubbcluj.ro}
\affiliation{Faculty of Physics, Babe\c{s}-Bolyai University,  Cluj-Napoca, 400084 Romania}
\affiliation{Romanian Institute of Science and Technology,  Cluj-Napoca, 400487 Romania}
\affiliation{Transylvanian Institute of Neuroscience,  Cluj-Napoca, 400157 Romania}

\date{\today} 

\begin{abstract} 
We introduce a continuous-time analog solver for MaxSAT, 
a quintessential class of NP-hard discrete optimization problems, 
where the task is to find a truth assignment for a set of Boolean 
variables satisfying the maximum number of given logical 
constraints. We show that the scaling of an invariant of the 
solver's dynamics, the escape rate, as function of the number 
of unsatisfied clauses can predict the global optimum value, 
often well before reaching the corresponding state. We demonstrate 
the performance of the solver on hard MaxSAT competition 
problems. We then consider the two-color Ramsey number $R(m,m)$ 
problem, translate it to SAT, and apply our algorithm to the 
still unknown $R(5,5)$. We find edge colorings without monochromatic 
5-cliques for complete graphs up to 42 vertices, while on 43 
vertices we find colorings with only two monochromatic 5-cliques, 
the best coloring found so far, supporting the conjecture that $R(5,5) = 43$.
\end{abstract}

\maketitle

Digital computing, or Turing's model of universal computing is 
currently the reigning computational paradigm. However, there are 
large classes of problems that are intractable  on digital computers, 
requiring resources (time, memory and/or hardware) for their solution 
that scale exponentially in the input size of the problem (NP-hard) \cite{MertensMoore}. 
Such problems, unfortunately, are abundant in sciences and 
engineering, for example, the ground-state problem of spin-glasses 
in statistical physics \cite{JPhysA_B82, STOC_I00}, the traveling 
salesman problem \cite{TSP}, protein folding \cite{prot}, bioinformatics 
\cite{bioinf}, medical imaging \cite{biomed1, biomed2}, 
scheduling \cite{GJ90}, design debugging, FPGA 
routing \cite{FPGA}, probabilistic reasoning \cite{ProbReas1, ProbReas2}, etc. 
As CMOS-based digital computing is reaching its limits 
\cite{Moore}, alternative approaches are being explored, such as 
analog computing and quantum computing. While
the latter seems promising, there are fundamental physics challenges 
that still need to be solved before it realizes its potential, leaving
analog computing as a currently exploitable option. Although it was 
explored in the '50s, it has been abandoned in the favor 
of the digital approach, due to the technical challenges it posed; 
for a historical survey on analog machines see \cite{Ulmann}.  By now, 
however, technology matured enough to control well
the physics at the small-scale and thus it makes worthwhile revisiting the 
analog branch of computing, at least at an application-specific level. 
For this reason, there has been an increasing effort dedicated recently 
to analog computing methods and devices specialized to address certain 
classes of problems \cite{Inagaki, Takata, McMahon, Marandi, Gert, 
Indiveri, Suman, Parihar, Emre, Zohar1,Zohar2,Ventra, Traversa, Adel, Gauthier, Stanley, Compiler}. 
However, there have been recent advances also in general purpose 
analog computing, see the reviews \cite{Pekka, MacLennan,Olivier0, 
Olivier1}, and in analog computability theory \cite{Olivier1, Olivier2, Kia}. 

Here we focus on fully analog (continuous-time) systems, in which both the 
state variables $\bm{s} = (s_1,\ldots,s_N)$ and the time variable $t$ are real 
numbers, $s_i \in \mathbb{R}$, $t \in \mathbb{R}$, updated continuously
by the algorithm (software), in form of a set of ordinary differential equations 
(ODEs) $d\bm{s}/dt = \bm{F}(\bm{s}(t),t)$, $t \in \mathbb{R}$,  \cite{Olivier1}. 
The process of computation is interpreted as the evolution of the trajectory 
(the solution to 
the ODEs) $\bm{s}(t) = \bm{\Psi}_t(\bm{s}_0)$, towards an attractive 
fixed-point state $\bm{s^*}$: $\lim_{t \to \infty}\bm{\Psi}_t(\bm{s}_0) = \bm{s^*}$, 
representing the answer/solution to the problem. Clearly, we want to 
find $\bm{s^*}$, and the challenge is to design $\bm{F}$  such that 
the solutions to the problem (when they exist) appear as attractive fixed points for the 
dynamics and no other, non-solution attractors exist that could trap the 
dynamics.  One such, first-principles based, continuous-time deterministic 
dynamical system (CTDS) has recently been proposed as an analog solver for Boolean satisfiability 
(SAT) in \cite{NatPhys_ET11}. 

In SAT we are given a set of $M$ logical clauses in conjunctive normal 
form (CNF), $C_1,C_2,\ldots,C_M$ over  Boolean variables $x_1,\ldots,x_N$, $x_i \in 
\{0,1\}$. Typically, one studies $k$-SAT problems where every clause 
involves $k$ literals (a literal is a variable or its negation). The task is to 
set the truth values of all the variables such that {\it all the clauses} 
evaluate to TRUE (``0''~=~FALSE, ``1''~=~TRUE). 
It is well known that $k$-SAT with $k \geq 3$  is NP-complete and thus 
any efficient solver for 3-SAT implies an efficient solver for all problems 
in the NP class (Cook-Levin theorem, 1971) \cite{STOC_C71,
GareyJohnson79}. The NP class is the set of 
all decision-type problems where one can check in polynomial time the 
correctness of a proposed solution (but finding such a solution can be 
exponentially costly). SAT has a very large number of applications in both 
science and industry, becoming a dominant back-end technology, lately. 
Applications include scheduling (crop rotation schedules, flight schedules), 
planning and automated reasoning (AI, robotics), electronic design automation, 
circuit design verification, bounded model checking for software/hardware 
systems (industrial-property verification), design of experiments, correlation 
clustering, coding theory, cryptography, drug design, etc. SAT is part of 
sub-problems in many domains such as test pattern generation, optimal control, 
protocol design (routing), image processing in medical diagnosis, electronic 
trading and e-auctions. For reviews see \cite{SATRev1, 
DiscApplMath_KS07} and the book \cite{SATBook}. 

The SAT solving system of ODEs proposed in \cite{NatPhys_ET11} was designed 
such that all SAT solutions appear as attractive fixed points for the dynamics 
while no other, non-solution attractors exist that could trap the dynamics. 
Note that for hard problems the dynamics of this solver becomes chaotic, 
showing that problem hardness and chaos \cite{SciRep_ET12,PRE_16} are related 
notions within this context, and thus chaos theory provides a novel set of 
tools to study computational complexity. When the SAT problem admits solutions, 
the chaos is necessarily transient 
\cite{LaiTel11,PhysRep_TL08}, as the trajectory eventually settles onto 
one of its attracting fixed points (a SAT solution). The CTDS was shown 
to solve hard SAT problems in polynomial time \cite{NatPhys_ET11}, but at the expense    
of auxiliary variables growing exponentially. In a hardware realization, this 
implies a trade-off between time and energy costs. However, since one 
can control/generate energy much better than time itself, this presents a 
viable option for time-critical applications. \cite{Xunzhao} proposes 
an analog circuit design for the CTDS from \cite{NatPhys_ET11}, showing 
a $10^4$-fold speedup (nanoseconds vs. milliseconds) on hard 3-SAT 
problems, when compared to the solution times by state-of-the-art digital SAT 
solvers (MiniSAT and variants \cite{minisat,SATsolvers}) on the latest digital processors. 

Here we propose a variant of the CTDS of \cite{NatPhys_ET11}, to solve 
MaxSAT problems. MaxSAT, or $k$-MaxSAT, has the same formulation 
as SAT (or $k$-SAT), but the task is to maximize the number of satisfied clauses 
(alternatively, to minimize the number of unsatisfied ones) and thus 
one cannot guarantee in polynomial time the optimality of the 
solution (unlike for SAT), for problems that do not admit full satisfiability. 
For this reason, MaxSAT is NP-hard.

The idea behind our approach is based on the observation that the 
operation of the CTDS SAT solver from \cite{NatPhys_ET11} does not assume 
full satisfiability and thus, even for SAT problems that do not admit full 
solution, the dynamics will still minimize the number of unsatisfied clauses. 
What we need to provide, however, is a method by which one can 
determine the likelihood of the optimality of the best solution found by 
analog time $t$, as function of $t$.  We achieve this heuristically, by 
analyzing the statistical properties of the chaotic behavior of the solver 
for hard problems, through one of its dynamical invariants, the escape rate 
\cite{LaiTel11,PhysRep_TL08}. Note that when the SAT problem admits 
no full solution (a true MaxSAT problem) there are no attractive fixed 
points, so the system is permanently chaotic. Defining the ``energy" of 
the system as the number of unsatisfied clauses (in a given instant), 
here we introduce the notion of ``energy"-dependent escape rate. 
The dependence of this measure on the number of unsatisfied constraints 
(energy) helps us to predict the energy level of the global optimum and to 
estimate the expected analog time needed by the solver to find it. 

We first perform a statistical analysis of the solver's performance on random 
3-MaxSAT problems and demonstrate that the solver works well for problems 
with good statistics on energy levels and thus for large problems. 
We then demonstrate the performance of the solver on very hard 
MaxSAT benchmark problems taken from recent MaxSAT competitions, 
including on problems that no competition solver could handle.  

Finally, we turn to the famous problem of Ramsey numbers 
\cite{GrahamSpencer1990, RamseyTheory} and we show 
how it can be translated into CNF SAT and then tackled with our solver. 
The Ramsey number $R(m,m)$ is the smallest order, complete graph such that 
no matter how we color its edges with two colors, we cannot avoid 
creating monochromatic cliques of order $m$. 
Thus, for complete graphs of order less than the 
Ramsey number, the coloring is a fully solvable SAT problem, whereas 
at the Ramsey number, the coloring problem becomes MaxSAT for 
the first time. The case of $m=5$ Ramsey number is still an open 
problem, only the bounds $43 \leq R(5,5) \leq 48$ are known 
\cite{RamseyNumbersSurvey, McKayRadz}. Finding $R(m,m)$ for a given
$m$ is very challenging because the search space is huge: there are 
$2^{N \choose 2}$ possible colorings of a complete graph on $N$ nodes. 
Thus, if $N=43$ is the Ramsey number for $m=5$, then the search space   
has $\approx 10^{271}$ possible colorings, impossible to search na\"ively.
Ramsey theory in general, deals with the unavoidable appearance of 
order in large sets of objects partitioned into few classes \cite{GrahamSpencer1990, RamseyTheory}.
It has deep implications virtually in all areas of mathematics, including
graph theory, combinatorics, set theory, logic, analysis and geometry 
\cite{Bollobas}. It has practical applications for example, in communications,
information retrieval and decision making \cite{Roberts1984}. 

Here we show how our algorithm finds good colorings (avoiding 
monochromatic $m$-cliques) for complete graphs of 
order less than the Ramsey number and then a prediction on the Ramsey 
number itself; for example, finding $R(4,4) = 18$ (a known result). 
For $m=5$ (equivalent to a $10$-SAT/MaxSAT problem) it finds good colorings for up to $N=42$ vertex complete graphs, whereas for $N=43$ it finds a coloring with only two monochromatic 5-cliques 
sitting on 6 nodes, the lowest energy coloring found so far, to our best 
knowledge. Given the efficiency of our algorithm to find solutions, this adds
further support to the conjecture that $R(5,5) = 43$. In all cases (competition and the Ramsey problems) 
we provide the solutions and the matrix of colorings in the Supplementary Information 
Section. We conclude with a brief discussion on analog solvers and their 
realization in hardware.

\section*{\large Results}

\section*{The MaxSAT problem}

Boolean satisfiability in conjunctive normal form (CNF) is a constraint 
satisfaction problem formulated on $N$ Boolean variables $x_i \in \{0,1\}$, 
$i = 1,\ldots,N$ and  $M$ clauses $C_1,\ldots,C_M$. A clause is the 
disjunction (OR operation) of a set of literals, a literal being either the 
normal ($x_i$) or the negated (NOT) form ($\overline{x}_i$) of a variable, 
an example clause being: $C_4 = (x_9\vee\overline{x}_{10}\vee x_{27})$. 
The task is to find an assignment for the variables such that all clauses are satisified, 
or alternatively, the conjunctive formula ${\cal F} = C_1\wedge \ldots 
\wedge C_M$  evaluates to 1 (TRUE). If all clauses contain exactly $k$ 
literals, the problem is $k$-SAT. For $k \geq 3$  this is an NP-complete 
decision problem \cite{STOC_C71}, meaning that a candidate solution is 
easily (poly-time) checked for satisfiability, but finding a solution can be 
hard (exp-time). Oftentimes, when studying the performance of 
algorithms over sets of randomly  chosen problems  the constraint 
density $\alpha = M/N$ is used as a rough statistical guide to problem 
hardness \cite{SciKS94, Science_MPZ02}.

Max-SAT is a more general version of SAT in that we must find an 
assignment that satisfies the maximum number of constraints (clearly, 
when the formula is satisfiable it is the same as SAT). We will define as 
the ``energy'' variable the number $E(\bm{x})$  of unsatisfied clauses 
given an assignment $\bm{x}$, and thus our task is to find an assignment 
of the variables corresponding to the global minimum of this energy function. 
For both SAT and MaxSAT, all known algorithms require exponentially 
many computational steps (in $N$)  in the worst case, to find a solution. 
However, unlike for SAT, checking the correctness for MaxSAT is as hard 
as finding the solution itself, thus making the problem harder than SAT 
(NP-hard).

\section*{A continuous-time dynamical system solver for SAT}

Here we briefly review the CTDS SAT solver introduced in \cite{NatPhys_ET11}  
and then modify it such as to be able to handle MaxSAT problems as well. 
The main strength of the solver is a one-to-one correspondence between 
the stable attractors of the dynamical system and the solutions of the SAT 
problem without introducing non-solution attractors. Starting the dynamics 
from almost all random initial conditions, it will keep searching until it finds 
a solution. For hard SAT formulas, the dynamics is transiently chaotic, 
revealing an interesting relation between chaos and problem hardness 
\cite{NatPhys_ET11,PRE_16}. However, when there is no solution satisfying all constraints, 
the global optimum is not a stable attractor anymore and we have to provide 
a method that can estimate the likelihood that the best solution found by 
analog time $t$ is the optimal MaxSAT solution. 

	To introduce the analog solver, we assign a variable $s_i = 2 x_i - 1$  
to every Boolean variable $x_i$  (when $x_i = 0$, $s_i = -1$  and when 
$x_i = 1$, $s_i = 1$), but allow $s_i$ to vary continuously in the $[-1,1]$ 
interval. The continuous dynamical system $\frac{d\bm{s}}{dt} = 
\dot{\bm{s}} = \bm{F}$ thus generates a trajectory confined to the hypercube 
${\cal H}_N = [-1,1]^N$. Clearly, the SAT solutions ${s^*}$  are all located 
in the corners of ${\cal H}_N$. To every clause $C_m$ (constraint) we 
associate the analog clause function $K_m(\bm{s}) = 2^{-k} \prod_{j=1}^N
(1-c_{mj}s_j)$, where $c_{mj} = 1$~(-1)  if variable $x_i$ appears in 
normal (negated) form in clause $C_m$,  and $c_{mj} = 0$ if it is missing 
(in either form) from $C_m$. The normalization $2^{-k}$ ensures that 
$K_m \in [0,1]$. One can easily check that $K_m = 0$ only in the corners 
of ${\cal H}_N$ and if and only if (iff) clause $C_m$ is satisfied. To define the 
dynamics of the system we introduce a ``potential energy'' function $V$  
that depends on the $K_m$-s such that $V=0$ iff all the clauses are 
satisfied, that is, $K_m = 0$, $\forall m = 1,\ldots,M$. A natural form for the 
potential energy is: 
\begin{equation}
V(\bm{s},\bm{a}) = \sum_{m=1}^{M}a_m K_m(\bm{s})^2 \;. \label{V1}
\end{equation}
Here the $a_m$ are time-dependent, positive weights,  $a_m > 0$, $\forall 
m = 1,\ldots,M$, $\forall t \geq 0$. If these weights were constants, the 
dynamics would easily get stuck in non-solution attractors. To prevent that, 
the dynamics of the auxiliary variables $a_m$ is coupled with the evolution 
of the clause functions $K_m$. The dynamics for the $\bm{s}$ variables is 
a simple gradient descent on the potential energy function $V$, the full 
system being:
\begin{empheq}[left=\empheqlbrace]{align}
&\frac{d \bm{s}}{dt} = \dot{\bm{s}} = - \nabla_s V \label{sdyn} \\[3mm]
&\frac{d \bm{a}}{dt} = \dot{\bm{a}} = \mathbb{K} \bm{a}, \;\;\mathbb{K} = 
\mbox{diag}(K_m)\;, \label{adyn}
\end{empheq}
where $\bm{a}$ is to be interpreted as a column vector with components 
$a_m$. Clearly, Eq. \eqref{adyn} preserves the positivity of the auxiliary 
variables at all times, since the analog clause functions stay non-negative 
at all times. According to \eqref{adyn} the auxiliary variables grow exponentially 
whenever the corresponding clause functions are not satisfied, however, 
once $K_m = 0$, $\dot{a}_m = 0$ and it stops growing. Eq. \eqref{adyn} 
ensures that whenever the dynamics would get stuck in a local, non-solution 
minimum of $V$, the exponential acceleration changes the shape of $V$ 
such as to eliminate that local minimum. This can be seen by first solving 
formally \eqref{adyn}: $a_m(t) = a_{m0} \exp\left(\int_{0}^t d\tau 
K_m(\bm{s}(\tau))\right)$, then inserting it into (1): $V = \sum_{m=1}^{M} 
a_{m0} e^{\int_{0}^t d\tau K_m} K_m^2$. Due to the exponentially growing 
weights, the changes in $V$ are dominated by the clause that was 
unsatisfied the longest. Keeping only that term in $V$ and inserting it 
into \eqref{sdyn}, it is easily seen that the dynamics drives the 
corresponding clause function towards zero exponentially fast, until another 
clause function takes over and this is repeated until all clauses are 
satisfied, for solvable SAT problems. 
In this sense, this is also a focused search dynamics \cite{Focused}. The 
properties and performance of this solver have been discussed in previous 
publications \cite{NatPhys_ET11,SciRep_ET12,PRE_16}. 
For hard (but satisfiable) SAT formulas the dynamics is transiently chaotic, 
but eventually all trajectories will converge to a solution. Since the dynamics 
is hyperbolic \cite{NatPhys_ET11}, the probability $p(t)$  of a trajectory not finding a 
solution by analog time $t$ decreases exponentially: $p(t) \sim e^{-\kappa t}$. The decay 
rate $\kappa$ is an invariant of transient chaos, called the {\it escape rate} 
\cite{PNAS_KT84,chaosbook}, and it well characterizes the hardness of the given SAT formula/instance. 
In \cite{SciRep_ET12} we demonstrated this on Sudoku puzzles (all Sudoku problems 
can easily be translated into SAT), showing that $\eta = - \log \kappa$ 
indeed provides a hardness measure that correlates well also with human 
ratings of puzzle hardness.

\section*{A continuous-time dynamical system solver for MaxSAT}

\begin{figure}[t!] 
\centering \includegraphics[width=0.9\textwidth]{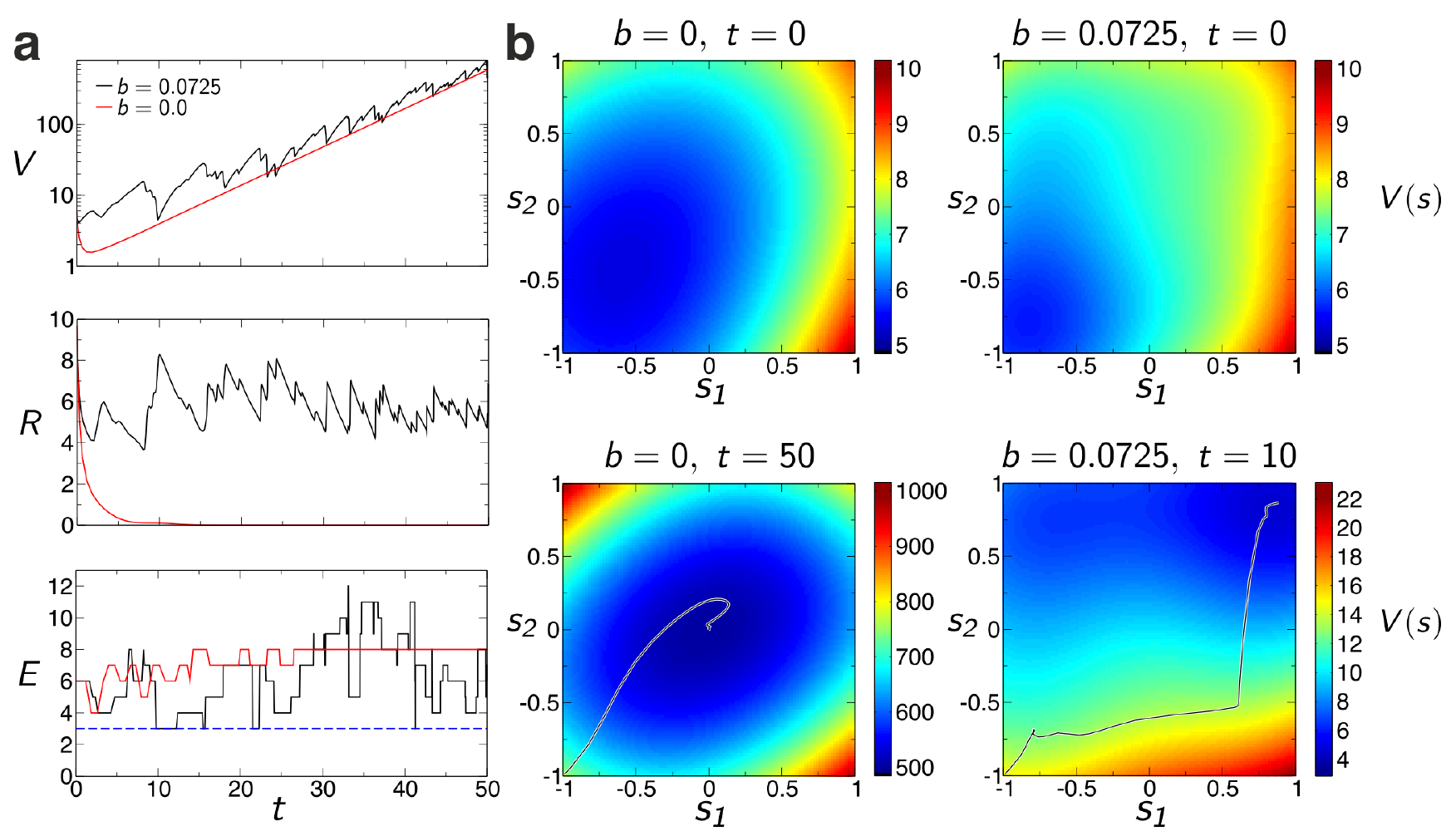}
\caption{{\bf MaxSAT solver dynamics.} The 3-MaxSAT formula used here has $N=10$, $M=80$ (clauses given in the Supplementary Table S1.) {\bf (a)} The potential $V$, the radius $R = \sqrt{\sum_i s_i^2}$, and the number of unsatisfied clauses (energy) $E$ as function of analog time $t$ for the  original dynamics corresponding to $b=0$ (red) 
and the modified dynamics with $b=0.0725$ (black). 
{\bf (b)} Colormaps of the potential $V(\bm{s}(t),\bm{a}(t))$ in the plane $(s_1,s_2)$. 
At a given time instant $t$  we fix all values $s_j(t), j=3,
\dots,N$ and $a_m(t),\forall m$ and change only $s_1,s_2$ in the 
$[-1,1]\times[-1,1]$ plane, showing the instantaneous 
potential energy landscape $V$ in this plane. The curves indicate the projection of 
the trajectory onto $(s_1,s_2)$ up to the indicated time $t$. In $t=0$,  $s_1=s_2=-1$. 
For $b=0$, the dynamics converges to $\bm{s}=0$, which is the centre of  a deep well 
in the potential landscape. For $b=0.0725$, the centre is not a minimum 
anymore and at time $t=10$ the orthant with minimal energy $E_{min}=3$ 
is found (the solution), shown as a blue dotted line in the $E(t)$ figure.} \label{fig1} 
 \end{figure}

Next we introduce a modified version of the above solver to solve the 
optimization type MaxSAT problem. In the dynamics presented in Eqs. 
(\ref{sdyn}-\ref{adyn}), 
if the global optimum $\bm{s^*}$  is not a solution with $V=0$, then  
$V$ will keep changing in time as function of the auxiliary variables. 
The dynamics is still biased to flow towards the orthants of the ${\cal H}_N$ 
hypercube with low energy, and as shown, in Fig.\ref{fig1}a, it will find the 
global optimum, but it will never actually halt there. Naturally, the question 
arises: How do we know when we have hit an optimal assignment? To 
tackle this question, we use a heuristic based on a statistical approach: 
we start many (relatively short) trajectories from random initial conditions, 
look for the lowest energy found by each trajectory and then exploit this 
statistic to help predict the lowest energy state and the time needed to 
get there by the solver. 

However, the dynamics (\ref{V1}-\ref{adyn}) cannot directly be applied 
for MaxSAT problems, one needs to modify the potential energy function, 
first. To see why, notice that the potential $V$ in the center of ${\cal H}_N$, 
in $\bm{s} = 0$, is always $V(\bm{0},\bm{a}) = 2^{-2k} \sum_{m=1}^M a_m$, 
because $K_m(\bm{0}) = 2^{-k}$, $\forall m$. On the other hand, in a corner 
$\bm{s'}$ of the hypercube, where $|s'_i| = 1$ $\forall i$, the value of each  
$K_m(\bm{s'})$ clause function is 0 if the clause is satisfied or 1 if it is unsatisfied, 
so the potential $V$ in a corner is just the sum of auxiliary variables 
corresponding to the unsatisfied clauses, i.e., $V(\bm{s'},\bm{a}) = 
\sum_{\{m: K_m \neq 0\}} a_m$.  Let $a$ be the average value 
of the auxiliary variables in a given time instance $t$,  
$a = \frac{1}{M} \sum_{m=1}^M a_m$. 
Thus $V(\bm{0},\bm{a}) = 2^{-2k} a M$  and $V(\bm{s'},\bm{a}) \simeq  
a E(\bm{s'})$, where $E(\bm{s'})$ is the number of unsatisfied clauses in 
$\bm{s'}$ (energy).  If $\bm{s'}$ is the global optimum and it is large enough 
(typically at large constraint densities $\alpha$),  the center of the hypercube 
may have a smaller potential energy value (due to the $2^{-2k}$ factor), than 
any of the corners of the hypercube, and it may become a stable attractor 
trapping the dynamics there. Figure \ref{fig1} shows an example of how this 
can happen on a small MaxSAT problem with $N=10$ variables and
$M=80$ clauses, given in the Supplementary Information (Table S1).  To prevent such 
trapping, we need to modify the potential 
energy function. We do this by adding a term $V'(\bm{s},\bm{a})$  to 
$V(\bm{s},\bm{a})$ such that it satisfies the following conditions: 1) it is 
symmetric in all $s_i$ so that there is no bias introduced in the search 
dynamics, 2) the energy in $\bm{s}=\bm{0}$ is always sufficiently large 
so that it never becomes an attractor 3) the added term does not modify 
the energy in the corners of the hypercube, and 4) similarly to the original 
dynamics, $\bm{s}$ always stays within the hypercube ${\cal H}_N$, which 
demands that all the partial derivatives  $\partial V'/\partial s_i$ vanish 
along the boundary of ${\cal H}_N$. We may imagine this added term in 
the form of a ``hat" function: it has a maximum at $\bm{s}=\bm{0}$  that 
keeps growing together with the time-dependent auxiliary variables (never 
to become permanently smaller than  the potential energy in the global 
optimum), but vanishing at the boundary surface of the hypercube. 
There are several possibilities for such terms, here we focus on one version 
that works well in simulations: 
\begin{equation}
V(\bm{s},\bm{a}) = \sum_{m=1}^{M}a_m K_m(\bm{s})^2 + 
b \alpha a \sum_{i=1}^ N \cos^2\left(\frac{\pi}{2}s_i\right), \label{V2}
\end{equation}
where $a$ is the average value of the auxiliary variables, $\alpha = M/N$ 
is the constraint density and $b$ is a constant factor tuning the strength 
of the last term to be always larger than the first, when chosen properly. 
The sum with the $\cos^2\left(\pi s_i/2\right)$ terms ensures 
the symmetric hat form, vanishing in the corners of ${\cal H}_N$. 
Note that the first term on the rhs of \eqref{V2} is never larger than 
$a M$. We now have $V(\bm{0},\bm{a}) = \left(2^{-2k} + b\right)aM$ 
and $b$ can be chosen such as to avoid the trapping phenomenon by 
the origin as described above, see also  Fig.\ref{fig1}b. 
To do that, we simply demand that 
the potential in the origin $V(\bm{0},\bm{a})$ keeps growing approximately
at the same rate as the potentials in the corners of the hypercube, never 
getting smaller than the potential in the global minimum (the smallest 
potential value in the corners). Thus, as long as $V(\bm{0},\bm{a}) \geq 
V(\bm{s'},\bm{a})$, where $\bm{s'}$ is some corner of the hypercube 
accessed by the dynamics, the dynamics will not get attracted by the 
origin of the hypercube. Since $V(\bm{s'},\bm{a}) \simeq a E(\bm{s'})$, 
this implies that $b \geq \frac{1}{M} E(\bm{s'}) - 2^{-2k}$, where 
$E(\bm{s'})/M$ is the fraction of unsatisfied clauses in $\bm{s'}$. Clearly, the $b$ value can be chosen arbitrarily large, however, if it is too large, then it forces the 
dynamics to keep running close to the surface of the hypercube, somewhat 
lowering its performance. In practice, an $E' = E(\bm{s'})$ is easily found 
by running a trajectory with a sufficiently large $b$ value for some short 
time, then resetting $b \geq \frac{E'}{M} - 2^{-2k}$. If chosen this way the search 
dynamics itself is not sensitive to this parameter $b$. The new dynamical system 
thus becomes:
\begin{empheq}[left=\empheqlbrace]{align}
&\dot{s}_i = -\frac{\partial V}{\partial s_i} = \sum_{m=1}^M 2a_m c_{mi} 
K_{mi}(\bm{s}) K_{m}(\bm{s}) + \frac{\pi}{2} b \alpha a \sin\left(\pi s_i\right)\,, 
\;\;\;\forall i=1,\ldots,N 
\label{sdyn1} \\
&\dot{a}_m = a_m K_m\,,\;\;\; \forall m=1,\ldots,m\; \label{adyn1}
\end{empheq}
where in Eq.~\ref{sdyn1} we used the notation $K_{mi}(\bm{s})=K_m(\bm{s})/(1-c_{mi}s_i)$.

Fig.~\ref{fig1} illustrates the difference between the two dynamics (see also 
 Fig. S1). While for $b = 0$ (original system) the dynamics converges 
rapidly to $\bm{s} = \bm{0}$ (seen, e.g., by monitoring the radius 
$R^2 = \sum_i s_i^2 \to 0$), the modified system with a properly chosen 
$b > 0$ continues the search. It finds an orthant with the minimum 
energy quite quickly (by $t = 10$), but it does not halt there, it 
continues the dynamics and returns to this minimum repeatedly 
(e.g., around $t \approx 16,22,41$). Fig. \ref{fig1}b shows the potential 
energy function landscape $V(\bm{s},\bm{a})$ in the  $(s_1,s_2)$ plane.  

\section*{An energy-dependent escape rate}  

The escape rate is an invariant measure of the dynamics introduced 
for characterizing transiently chaotic systems \cite{PNAS_KT84,chaosbook}. In a transiently chaotic 
system the asymptotic dynamics is not chaotic, but, for example, settles 
onto a simple attractor, or escapes to infinity (in open systems), however, 
the non-asymptotic dynamics is chaotic, usually governed by a chaotic 
repeller. It is well known that for hyperbolic, transiently chaotic dynamical 
systems the probability of a randomly started trajectory not converging to 
an attractor by time $t$ (i.e., not finding a SAT solution in our case) 
decreases exponentially in time: $p(t) \sim e^{-\kappa t}$, where $\kappa$ 
is the escape rate \cite{LaiTel11,PhysRep_TL08,PNAS_KT84}. 
\begin{figure}[htbp] 
\centering\includegraphics[width=0.90\textwidth]{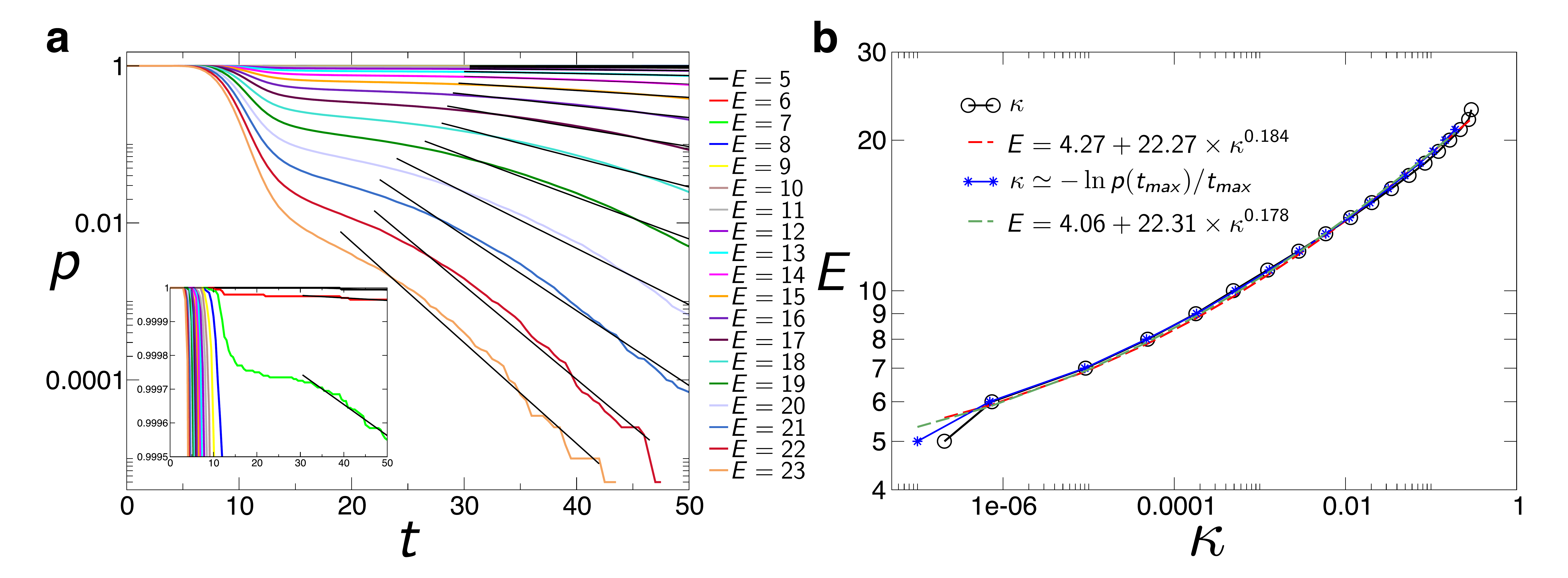}
\caption{{\bf Energy dependent escape rate.} {\bf (a)} The $p(E,\,t)$ 
distributions for a hard, benchmark MaxSAT competition problem with 
$N=250$, $M=1000$ ($\alpha=4.0$) (source: \cite{benchmark14,benchmark16}, also the first row in table of Fig. \ref{table}). We obtain $E_{min}=5$ with our algorithm after running $\Gamma =2\times 10^{5}$ trajectories with $b=0.002375$. The escape rates are obtained 
from fitting $p\sim e^{-\kappa t}$ onto the last section of the distributions 
(black lines). {\bf (b)} The energy $E$ vs the escape rate $\kappa$ using the 
values obtained from fitting shown in a) (black)  and using the rough 
estimation for the escape rates $\kappa(E)\simeq-\ln (p(E,\,t_{max}))/t_{max}$ 
(blue). This estimation is convenient, as it is much easier to automate in the algorithm
than the fitting procedure itself (see Methods). The dashed line curves show the 
fitting of Eq. \eqref{plfit}. Both curves result in $E_0\in (4,5]$ thus predicting 
the global optimum $E^{pred}_{min}=5$.} \label{fig2} 
\end{figure}
The escape rate can also be interpreted as the 
inverse of the average lifetime $\tau$ of trajectories $\kappa = 1/\tau$. 
For permanently chaotic systems, such as our MaxSAT solver, however, 
this definition does not work, as there is no simple asymptotic attractor 
in the dynamics and the system is closed. To be able to use a similar notion 
also for MaxSAT, we use a thresholding on the energy of the visited states. 
More precisely, we monitor the probability $p(E,\,t)$ that a trajectory has 
not yet found an orthant of energy smaller than $E$ by analog time $t$. Here
$E$ acts as a parameter of the distribution. 
This can be measured by starting many trajectories from random initial conditions and monitoring the fraction of those that have not yet found a state with an energy less than $E$ by analog time $t$. In Fig.\ref{fig2}a we show these distributions for different $E$ values for a MaxSAT problem. For large $E$, all trajectories almost immediately find orthants with fewer unsatisfied 
clauses, but for lower $E$ values the distributions decay exponentially. 
We call their decay rates energy-dependent escape rates $\kappa(E)$. 
Naturally, if an energy level does not exist in the system (e.g., for 
$E < E_{min}$), the escape rate for that energy level is meaningless 
(extrapolates to zero or a negative number).  
This suggests that the $\kappa(E)$ dependence could be used to predict 
where this minimum energy is reached. However, to capture this energy 
limit, it is more convenient to plot the $E(\kappa)$  function, instead 
(see Fig.\ref{fig2}b). From extensive simulations, we observe a power-law 
behavior with an intercept $E_0$:
\begin{equation}
E = E_0 + c \kappa^{\beta}\;. \label{plfit}
\end{equation}
Since $E_0$ is not an integer in general, we have $E_{min} = 
\lfloor E_0 \rfloor+1$. This observation is at the basis of our method to 
predict the global energy minimum for MaxSAT.

\section*{Procedure for predicting the global minimum}

Here we describe the algorithm that also provides the
halting criterion for the system (\ref{sdyn1}-\ref{adyn1}) searching
for solutions, with   details of the computations presented in the Methods section 
along with an algorithm flowchart shown in Fig. S2.
The exponentially decaying nature of the $p(E,\,t)$ distributions implies 
that sooner or later every trajectory will eventually visit the orthant with 
the lowest energy. Nevertheless, instead of leaving one trajectory to run 
for a very long time, it is more efficient to start many shorter 
trajectories from random initial conditions while tracking the lowest energy 
reached by each trajectory (see Fig. S3 and the
Discussion section). This also generates good statistics for $p(E,\,t)$, 
and for obtaining the properties of the chaotic dynamics that are then 
exploited along with \eqref{plfit} to predict the value of the global minimum 
and to decide on the additional number of trajectories needed to find 
a lower energy state with high probability. 

The basic step of the algorithm is to run a trajectory $\omega$ from 
a random initial condition up to a given time $t_{max}$ and record the lowest energy found by this particular trajectory, 
denoted by $E_s(\omega)$. Let $\Gamma$ denote the total number 
of trajectories run so far, ${\cal T}$ the set of these trajectories (thus 
$\Gamma = |{\cal T}|$), and $\overline{E}(\Gamma) = \min_{\omega 
\in {\cal T}} E_s(\omega)$ be the lowest energy found by all these 
trajectories. Using statistical methods and the relation between energy and 
escape rate $\kappa(E)$ (shown in \eqref{plfit}), the algorithm repeatedly predicts (as $\Gamma$
grows) the expected number of trajectories we need to run in total to find 
the lower energy value $\overline{E}-1$, i.e., $\Gamma^{pred}(\overline{E}-1)$ and the global minimum energy $E^{pred}_{min}$. We then 
monitor $E^{pred}_{min}$ for saturation and once the saturation criterion 
is reached, it outputs a decision $E^{dec}_{min}$, representing the 
final energy value predicted by the algorithm as the global minimum.
If this energy value has actually already been attained (found at least one
assignment for it), the algorithm outputs the corresponding assignment(s). 
If it did not attain it then it keeps running until finds such an assignment 
or reaches the preset maximum limit $\Gamma_{max}$ on the number 
of runs. In the latter case it outputs the lowest energy value attained and 
the corresponding assignment(s) and the consistency status of the 
predicted value. In the Methods section we provide the description 
of the algorithm. 

\section*{Performance on random 3-MaxSAT problems}

We first tested our algorithm and its prediction power on a large set 
(in total 4000) of random 3-MaxSAT problems with $N = 30, 50, 100$ 
variables and constraint densities $\alpha =8,10$. (In 3-SAT the SAT-UNSAT 
transition is around $\alpha \simeq 4.267$ 
\cite{MezardZecchina}). We compare our results with the 
true minimum values ($E_{min}$) provided by the exact algorithm MaxSATZ \cite{MaxSatzAAAI,MaxSatzJAIR}. 
\begin{figure}[htbp] 
\centering \includegraphics[width=0.90\textwidth]{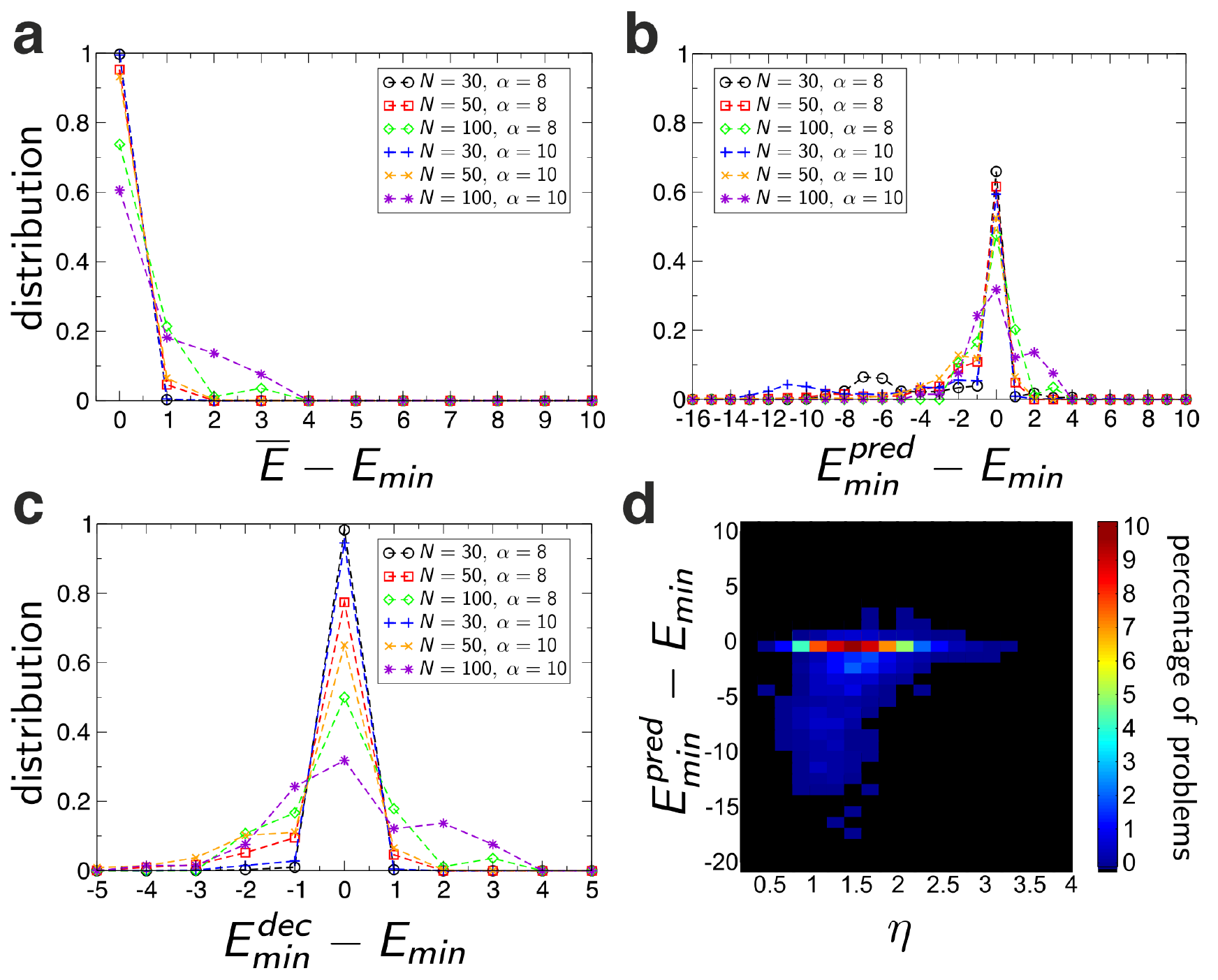}
\caption{{\bf Algorithm statistics over random 3-MaxSAT problems.} Distribution of differences between the real global minimum $E_{min}$ obtained with the exact algorithm MaxSatz and {\bf (a)} the smallest energy found by the algorithm  $\overline{E}$, {\bf (b)} the predicted minimum value $E_{min}^{pred}$ and {\bf (c)} the final decision of the algorithm $E_{min}^{dec}$ shown for problems with different $N$ and $\alpha$ values (see legends). {\bf (d)} The percentage of instances indicated by color (see color bar) for different values of the error $E_{min}^{pred}-E_{min}$ and  hardness $\eta$. Most instances are in the $E_{min}^{pred}-E_{min}=0$ row indicating correct prediction. Large errors  occur mainly at smaller $\eta$ values, and are dominantly negative.} \label{fig3} 
\end{figure}
In Fig. \ref{fig3} we compare the lowest energy found by the algorithm $\overline{E}$, 
the predicted minimum $E^{pred}_{min}$ and the final decision by the algorithm 
$E^{dec}_{min}$ with the true optimum $E_{min}$, by showing the distribution 
of differences across many random problem instances. We apply $t_{max}=25$ and limit the number 
of trajectories to $\Gamma_{max} = 150\,000$, after which we stop the algorithm 
even if the prediction and the decision are not final. For that reason, it is 
expected that the performance of the algorithm decreases as $N$ increases, 
(e.g., at $N=100$), we would need to run more trajectories to obtain the 
same performance. Nevertheless, the results show that all three distributions 
have a large peak at $0$. Naturally, the most errors occur in the prediction phase, but 
many of these can be significantly reduced through simple decision rules 
(see Methods),  because they occur most of the time in easy/small problems, 
where the statistics is insufficient (e.g., too few points since there are only few energy values).
 \begin{figure*}[htbp] \begin{center}
\includegraphics[width=0.90\textwidth]{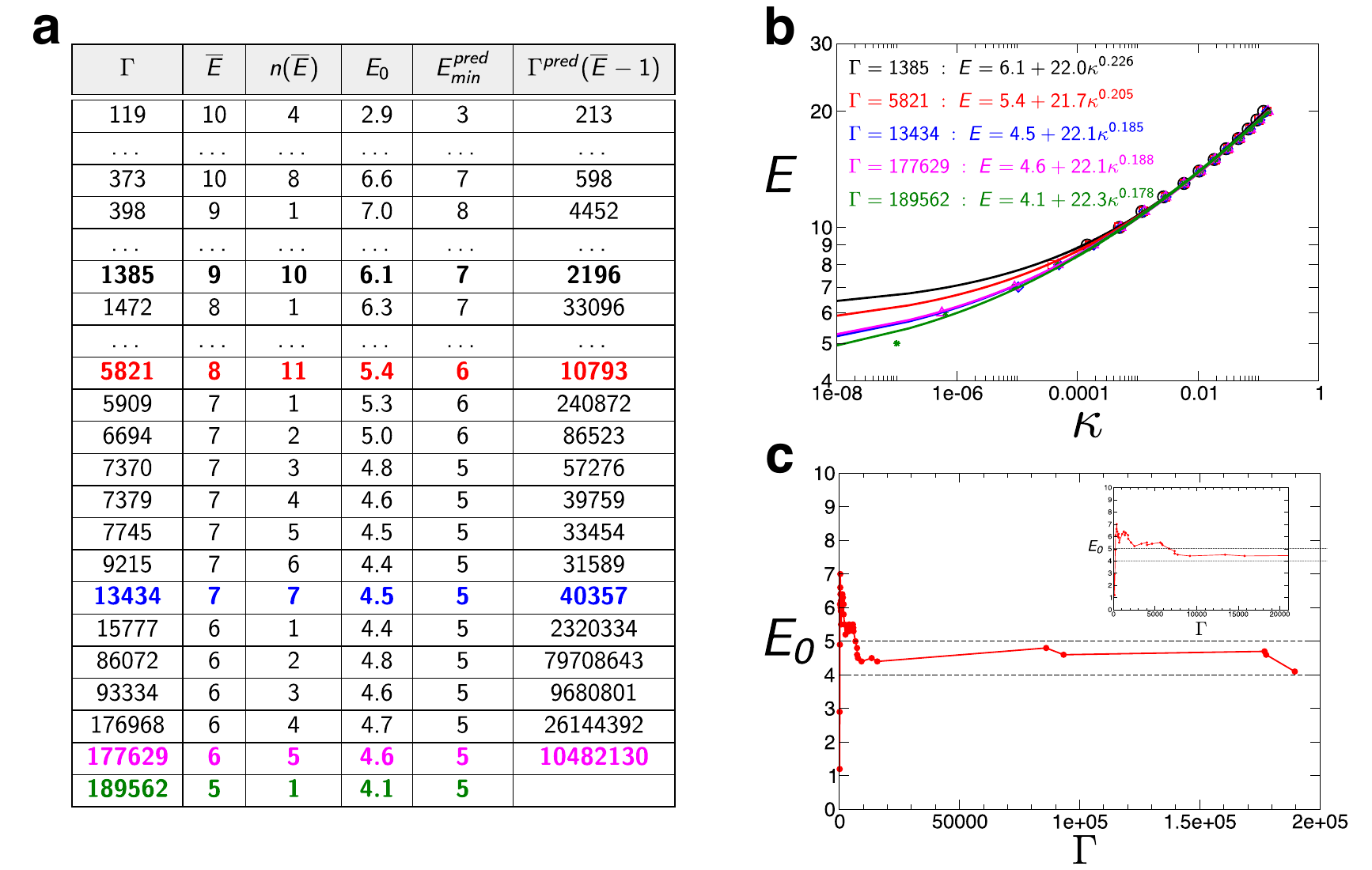}
\caption{{\bf Performance of the algorithm illustrated on  a hard benchmark problem.} 
We use the same problem as in Fig. \ref{fig2}.
{\bf (a)} $\Gamma$ is the number of trajectories, $\overline{E}$ the lowest energy found until that 
point, $n(\overline{E})$ is the number of times this energy has been found, $E_0$ is the 
parameter obtained from fitting of Eq. (\ref{plfit}), $E^{pred}_{min}$ and 
estimating $\Gamma^{pred}(\overline{E}-1)$. The algorithm estimates the escape rate and performs a prediction at each $\Gamma$ shown in the table and for the colored lines  we show the fitting curves in {\bf (b)}. {\bf (c)} The relevant parameter  $E_0$ is shown as function of $\Gamma$. 
While it wildly fluctuates at the beginning when the statistics is small, it remains in the 
$E_0\in[4,5)$ interval, convincingly predicting $E^{pred}_{min}=5$ already after 
$\Gamma=7\,000$ up until the point that it finds this energy at $\Gamma=189\,562$. 
At this point it could be expected that we do not have a good estimate for $\kappa(5)$ 
because it has been found only once ($n(5)=1$), nevertheless the estimation 
$E^{pred}_{min}$ remains consistently the same, convincing our algorithm to 
accept $E^{dec}_{min}=5$ and stop.
} \label{fig4} 
\vspace*{-0.5cm} \end{center} \end{figure*}  
To show how the error in prediction depends on the hardness of problems, 
we studied the correlation between the error $E^{pred}_{min} - E_{min}$ and 
the hardness measure applicable to individual instances $\eta = - \ln \kappa / \ln N$ 
(for the origin and definition of this hardness measure see \cite{SciRep_ET12}). 
In Fig. \ref{fig3}d we show the distribution of these values (also see 
 Fig. S4).  Interestingly, larger errors occur mainly at the easiest problems with $\eta < 2$. 
Calculating the Pearson correlation coefficient between  $| E^{pred}_{min} - E_{min}| $ 
and $\eta$  (excluding instances where the prediction is correct) we obtain a 
clear indication that often smaller $\eta$ (thus for easier problems) generates larger errors. Positive errors are 
much smaller and are shifted towards harder problems. There are somewhat more negative errors, which means that the algorithm consistently predicts a slightly lower energy value than the optimum, which is good, since this way we have an increased assurance that the algorithm has found the optimum state. In Fig. S4b we show the correlation coefficients calculated separately for problems with different  $N$, $\alpha$ settings.

\section*{Performance evaluation on hard MaxSAT competition problems}

\begin{figure}[htbp] 
\centering \includegraphics[width=\textwidth]{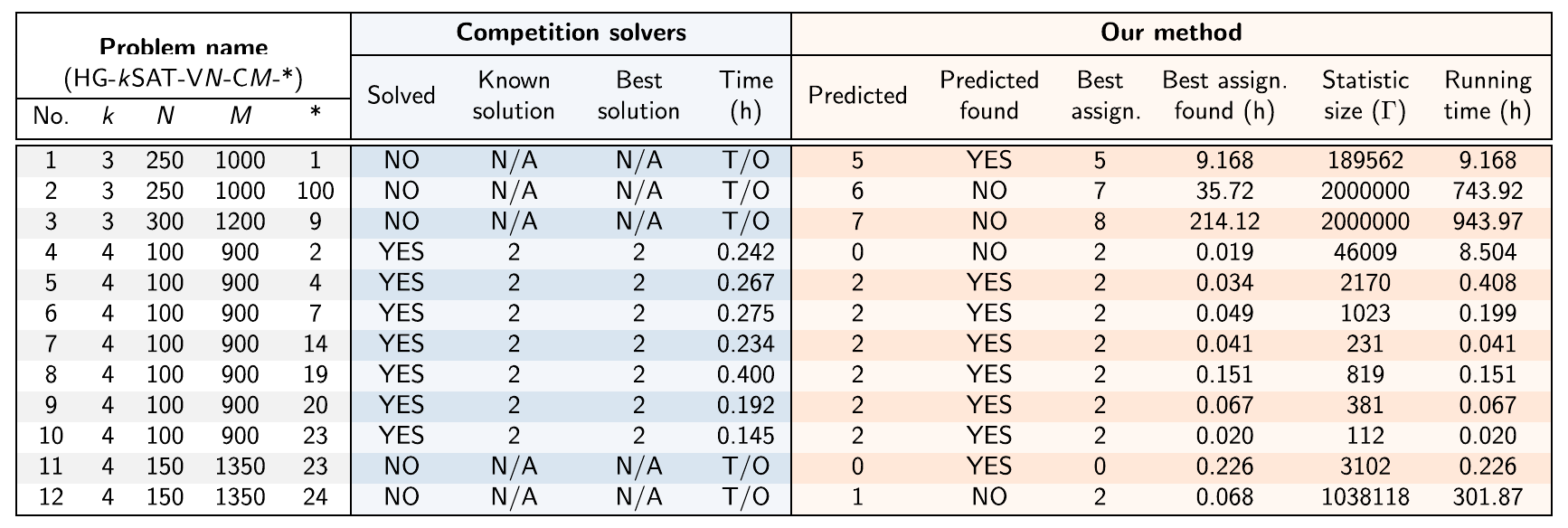}
\caption{{\bf Algorithm performance on competition MaxSAT problems.} The SAT instances can be downloaded from \cite{benchmark14,benchmark16}. The best solutions found by our algorithm are given in the SI.
 } \label{table} 
\end{figure}

We next present the performance of our solver on very hard MaxSAT 
problems taken from competitions, listed on the site \cite{benchmark14,benchmark16}. 
For illustration purposes, here we focus on an extremely hard competition 
problem instance, called {\tt HG-3SAT-V250-C1000-1.cnf}, which was 
reposted (for several years) with $N=250$ variables and $M = 1000$ clauses, 
shown as problem No. 1 in the table of Fig. \ref{table}. This problem was also used 
in Fig. \ref{fig2}. No competition algorithm could solve this problem, or 
even predict the minimum energy value within the allotted time (30 mins). 
We ran our algorithm on a regular 2012 iMac 21.5, 3.1GHz, Intel Core i7 
computer and it predicted the lowest energy of 5 (unsatisfied clauses), after 
  21min 24sec  of running time and 
produced an assignment for it after   9.168h  of running time. 
The minimum energy prediction was achieved already after $\Gamma = 7000$ trajectories, 
whereas an assignment with this minimum energy took a total of  
$\Gamma = 189\,562$ trajectories to run.  The minimum energy assignment corresponding to $E^{dec}_{min} = 5$ found by our 
algorithm is provided in the Supplementary Information section, Table S2. (The problem itself can be downloaded from the competition site \cite{benchmark16}). 
We ran the complete and exact algorithm, MaxSatz \cite{MaxSatzAAAI,MaxSatzJAIR} 
for more than 5 weeks on this problem and the smallest energy it found was $E = 9$! 
The details of how the algorithm performs are shown in Fig. \ref{fig4}.
Similar figures for other hard problems such as for a 4-SAT problem and a spin-glass problem are shown in Figs. S5, S6.

 The table of Fig. \ref{table} shows the performance of the algorithm on several 
hard competition MaxSAT problems taken from the same source (notice, 
9 out of 12 problems are 4-MaxSAT). Instances No. 1-3, 11, 12 were not solved 
by any competition solver, but our solver made a low energy prediction for all of them and 
for problems 1 and 11 it also did find an assignment corresponding to the predicted 
energy value. For No. 2, 3 and 12, it did not find a corresponding assignment, but, 
consistently finds an energy value that is only one higher (i.e., for 7 instead of 6, for 8 instead
of 7 and 2 instead of 1). The reason for this is that the solver, by the nature
of the fitting, sometimes predicts a lower energy value than the correct minimum one, as discussed previously.
Note that the corresponding best assignment is found relatively early (even
if the algorithm was run much longer, hoping to find a realization for the 
predicted value). Problems No. 4-10 were also solved by the competition
solvers, achieving the best known solution (of 2 unsatisfied clauses in all cases),
which was also found by our algorithm. In one instance, that of problem 4, 
it predicted an energy value of 0 (consistently with the observation of predicting
lower values than than the minimum), which, of course it could not find, but it did
find an assignment for the correct value of 2. In all these cases it found 
assignments faster than the competition solvers, often nearly ten times
faster. In the Supplementary Information section we provide the minimum energy 
values and solution assignments for all the problems presented in Fig.
\ref{table}.

\section*{Application to Ramsey numbers}

Next we further demonstrate the ability of our analog algorithm to solve 
exceptionally hard MaxSAT problems. The problem of Ramsey numbers 
or Ramsey theory is a very active area in mathematics with applications 
to virtually every field of mathematics \cite{Bollobas, RamseyTheory}.  
Although it has several variants, in the standard, two-color Ramsey number 
problem we have to find the order for the smallest complete graph for which 
no matter how we color its edges with two colors (red and blue), we cannot 
avoid creating a monochromatic $m$-clique. The number of nodes for the 
smallest such complete graph is denoted by $R(m,m)$. The more popular 
formulation for $m=3$ is going back to Paul Erd\H{o}s: minimum how many people 
$R(3,3)$  should we invite to a party to make sure that there are either 3 
people who mutually all know each other or 3 people who mutually do not 
know each other? (In this case one edge color corresponds to 
people knowing each other and the other to not knowing each other). The 
proof that $R(3,3)=6$ is trivial. 
For $m=4$  the answer is $R(4,4) = 18$ and it is harder to prove \cite{GG}. 
The $m=5$ case is still open, only the bounds $43 \leq R(5,5) \leq 48$ 
are known \cite{RamseyNumbersSurvey}. The best lower bound of $43$ 
was first found in 1989 by Exoo \cite{Ex1}, and the upper bound was only recently 
reduced from 49 \cite{McKayRadz} to 48 by Angeltveit and McKay \cite{R48}. 
Using various heuristic methods, including simulated annealing, tabu search and genetic 
algorithms, researchers have found in total 656 solutions (328 graphs and 
their complements) for the complete graph on 42 nodes \cite{McKayRadz}. 
It has been conjectured by McKay and Radziszowski \cite{McKayRadz} 
that there are no other solutions for $N=42$. 
Starting from these solutions they searched for a 5-clique-free coloring in 43 and
as no solution was found,  McKay, Radziszowski and Exoo made the strong 
conjecture that $R(5,5) = 43$ \cite{McKayRadz}. Other variants of Ramsey 
problems include specifying different clique orders for different colors and/or 
using more than two colors \cite{RamseyTheory}.

Ramsey number problems are very challenging because the search space is 
huge: it grows as $2^{N(N-1)/2} = O(2^{N^2})$ with the 
order $N$ of the complete graph to be colored.  To tackle  Ramsey number 
problems with our algorithm, we first transform them into $k$-SAT: 
every edge $i$ ($i=1,\ldots, N(N-1)/2$) to be colored is represented 
by a Boolean variable $x_i$ (with $x_i \in \{0,1\}$, 1 = blue, 0 = red). 
A clique of size $m$ has $m(m-1)/2$ edges. We are satisfied with a 
coloring (a solution) when no $m$-clique is monochromatic, in other 
words, every $m$-clique with set of edges $\left\{i_1,\ldots,i_{m(m-1)/2}\right\}$ 
must have both colors, expressed as the statement formed by the 
conjunction of the two clauses 
\begin{equation}
\left( x_{i_1}\vee \ldots \vee x_{i_{m(m-1)/2}} \right)\wedge
\left( \overline{x}_{i_1}\vee \ldots \vee \overline{x}_{i_{m(m-1)/2}}\right) 
\label{RSAT}
\end{equation}
being true. This means that for {\it every} $m$-clique we have two 
clauses and thus there are a total of $2 \binom{N}{m}$ clauses to satisfy. 
Since the number of clauses ($O(N^m)$)  for $m \geq 2$ grows faster 
in $N$ than the number of variables $N$, there will be a lowest $N$ 
value corresponding to UNSAT, which is the sought $R(m,m)$ Ramsey 
number. Thus, for $m=3$ we have a 3-SAT problem, for $m=4$ a 
6-SAT problem and for $m=5$ a 10-SAT problem. For graphs with $N=42$
nodes the number of clauses is $1\,701\,336$ and the search space has 
over $2^{861} \simeq 1.5 \times 10^{259}$ graphs! If we were to compute 
the familiar constraint density $\alpha$, it would be $\alpha = 
2 \binom{42}{5}/(21\times 42) = 1976$, indeed above the SAT/UNSAT 
transition point for random 10-SAT, which is estimated 
to be $\alpha_s{\big|}_{\mbox{\footnotesize 10-SAT}}\simeq 707$ \cite{10SAT}. 
\begin{figure}[t!] 
\centering \includegraphics[width=0.7\textwidth]{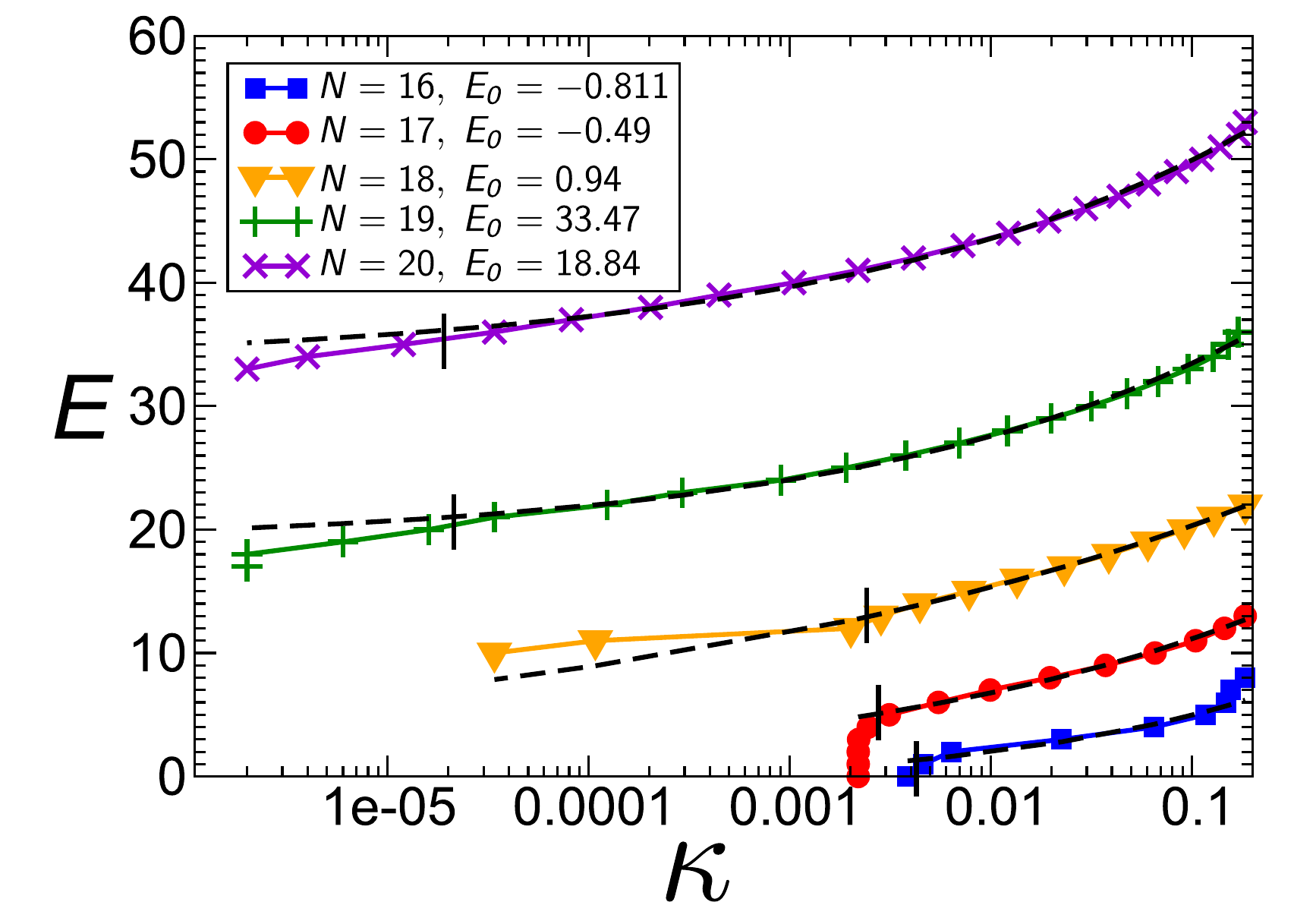}
\caption{{\bf Finding the Ramsey number $R(4,4)$.} The $E(\kappa)$ relationship
for the 6-SAT problems corresponding to the $K_{N}$ complete graph colorings
with two colors. $E_0$ is the extrapolated value based on the fit from 
Eq. \eqref{plfit} (dashed lines). The long vertical bars indicate the lower end of the
fitting range. Note that for $N = 16,17$, $E_0$ is a negative
value indicating full colorability (the corresponding 6-SAT problem is fully satisfiable), 
whereas for $N \geq 18$, $E_0>0$, and thus the 6-SAT problem becomes MaxSAT. } \label{fig6} 
\end{figure}

Applying our algorithm for the $m=4$  Ramsey problem, we can easily find 
coloring solutions for $N \leq 17$,  while for $N=18$ it
predicts that there is no solution, indeed confirming that $R(4,4) = 18$. 
This is seen from the plot of $E$ vs $\kappa$  in Fig \ref{fig6}. 
For $N \leq 17$  the smooth portion of the curve fitted by \eqref{plfit} 
suddenly cuts off, $\kappa$ being the same for all energy values lower than
a threshold value, meaning that after reaching a state corresponding to the 
threshold energy level, the solution (i.e., $E = 0$) is immediately found. 
This is simply due to the fact that the behavior \eqref{plfit} is a statistical 
average behavior characteristic of the chaotic trajectory, from the neighborhood of the chaotic repeller of the dynamics and away from the region 
in which the solution resides. However, once the trajectory enters the basin 
of attraction and nears the solution, the dynamics becomes simple, non-chaotic and runs into the solution, reflected by the sudden drop in energy.  
This is not due to statistical errors, because the curve remains 
consistent when plotting it using $10^3,10^4$ or $10^5$ initial conditions 
(the figure shows $10^5$ initial conditions).

\begin{figure}[t!] 
\centering \includegraphics[width=1.00\textwidth]{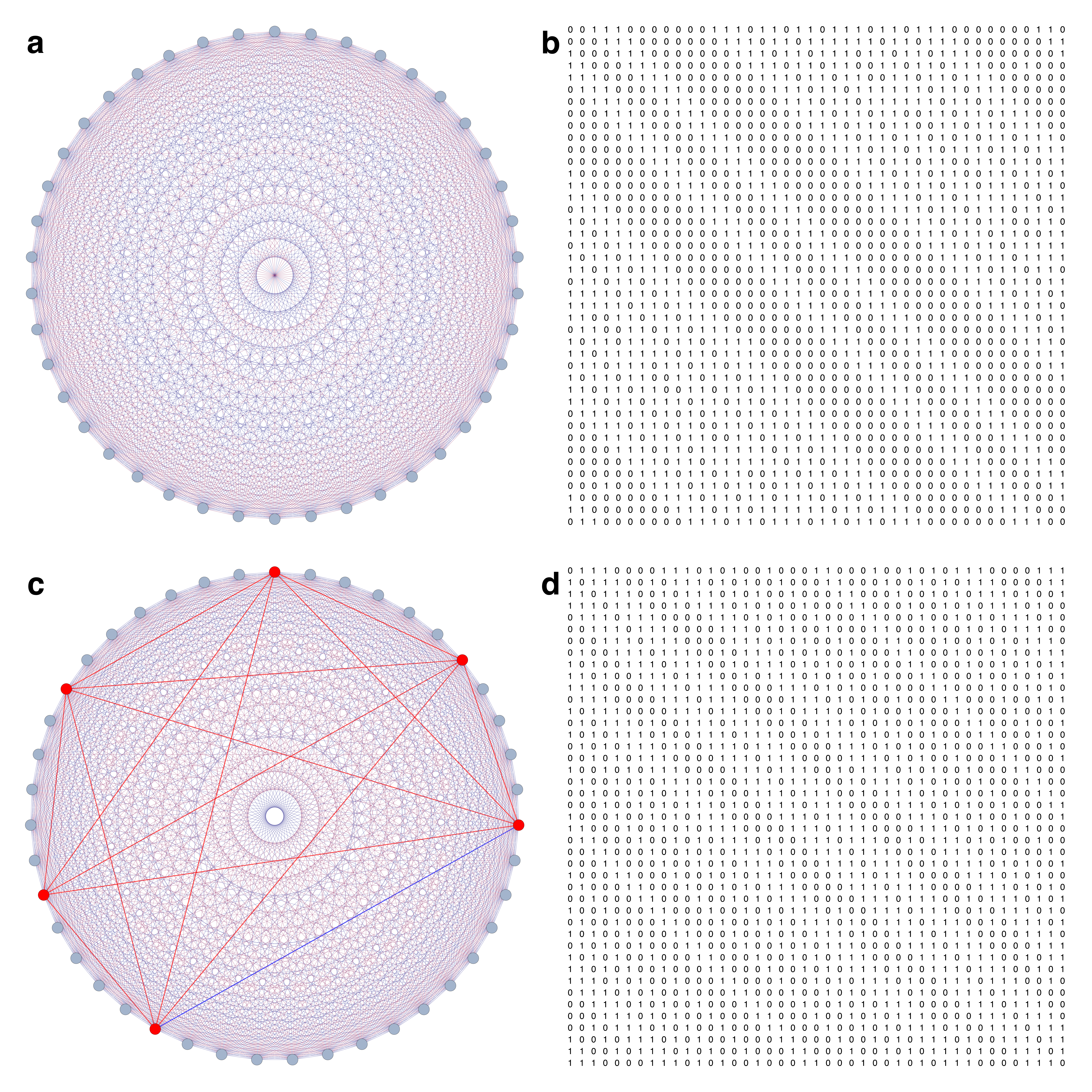}
\caption{{\bf Colorings for the $R(5,5)$ Ramsey number problem.} {\bf (a)} A coloring of the complete graph on $N=42$ nodes that avoids monochromatic 5-cliques. {\bf (b)}  The adjacency matrix corresponding to the coloring in {\bf (a)}.
{\bf (c)} The best coloring of the complete graph on $N=43$ nodes containing
only 2 monochromatic (red) 5-cliques, sitting on 6 nodes (highlighted with thicker edges). {\bf (d)}  
The adjacency matrix corresponding to the coloring in {\bf (c)}. } \label{fig7} 
\end{figure}

Searching for the value of $R(5,5)$ one can relatively easily find coloring 
solutions without 5-cliques up to $N=35$ for 
which the number of variables is $595$ and the number of clauses $649\,264$, 
already huge for a 10-SAT problem for other types of SAT and MaxSAT solvers. To find solutions faster for
$N \geq 36$, however, we employ a strategy based on circulant 
matrices \cite{Cyclic} helping us find solutions (proper colorings) up to and 
including $N=42$ in a relatively short time (on the order of hours), which we 
describe next.  

Kalbfleisch \cite{Cyclic} argued that there should be coloring solutions of
complete graphs for the Ramsey problem that can be described with a 
{\it circulant form} adjacency matrix (e.g., all red edges are 1-s, blue edges are 
0-s in this matrix), or matrices that are close to such a circulant form. 
This was then later applied by many authors in improving bounds for Ramsey 
numbers and finding corresponding colorings. Although there is no formal proof
of this statement, one expects this to be true also from the SAT formulation of the
Ramsey problem. In the SAT formulation, the clauses have a very high degree of
symmetry: all variables participate in the same way \eqref{RSAT} in all 
the clauses, which run over all the possible $m$-cliques.
This observation on symmetry can be exploited, allowing us to do part of the 
search in a much smaller space than the original space, where all the variables 
could in principle change independently from one another. More precisely,
we first define a MaxSAT problem which has only $N-1$ variables (instead of the
full $N(N-1)/2$) by choosing, 
e.g., those associated with the links of the first node: $x_1 = a_{1,2}, x_2 = 
a_{1,3}, \ldots, x_{N-1} = a_{1,N}$ as problem variables  
(here $a_{i,j}$ denotes the adjacency matrix) and defining the variables of the 
links of the other nodes  by the circular permutation of this vector $\bm{x}$, 
to obtain a circulant matrix (e.g., $a_{2,3} = a_{1,2}$,  $a_{i,j} = a_{i-1,j-1}$). 
Taking the MaxSAT form of the Ramsey problem we replace the variable of 
a link $a_{i,j}$ with $x_{j-i}$, thus reducing the number of independent 
variables from $N(N-1)/2$ 
to $N-1$. The number of clauses will also be reduced, because we can now
eliminate the repeated ones. This way we obtain a much smaller  MaxSAT 
problem, on which we apply our solver and starting from random initial conditions
we search for low-energy states, which are relatively easily found. We save the 
$\bm{x}$ vectors (the Boolean values) corresponding to such low-energy 
circulant matrix states.  For $N=42$ we have found circulant type 
matrices having only 6, 14, 20, 26 etc.  monochromatic 5-cliques, 
indicating that they may already be close to a solution. After saving these circulant matrix
states (with small number of monochromatic 5-cliques) we return to the original 10-SAT 
problem (with $N(N-1)/2$ variables, without the symmetry constraint), 
and start a new trajectory from the corner of the hypercube corresponding to the saved
matrix state, but now without symmetry restriction. 
This places the trajectories relatively close to the solution and a proper coloring 
can be found in hours even for $N=42$, (see Fig. \ref{fig7}a, b, and Table S15 
for an easily readable list of edge colorings), for which other 
heuristic algorithms take many days of computational time \cite{McKayRadz}, 
even with the circulant matrix strategy. 
Applying the same strategy for $N=43$ we did not find any complete coloring 
solutions, however, we did find a coloring that creates only {\it two} (out of $962\,598$ possible) 
monochromatic 5-cliques, see Fig. \ref{fig7}c, d, and the specific coloring 
provided in the Supplementary Information Table S16. This adds further 
support to the conjecture that $R(5,5) = 43$.

\section*{Discussion}

In summary, here we presented a novel, continuous-time dynamical system approach 
to solve a quintessential discrete optimization problem, MaxSAT. The solver is based
on a deterministic set of ordinary differential equations and a heuristic method that is
used to predict the likelihood that the optimal solution has been found by analog time 
$t$. The prediction part of the algorithm uses the statistics of the ensemble of trajectories 
started from random initial conditions, by introducing the notion of energy-dependent 
escape rate and extrapolating this dependence to predict both the minimum
energy value (lowest number of unsatisfied clauses) and the expected time needed by the
algorithm to reach that value. The statistical analysis of the ensemble of trajectories
presented here is very simple; it is quite possible that more sophisticated, extreme value 
statistical methods can be used to better predict minima values and time lengths.  
Due to its general character, the presented approach is generalizable to other 
optimization problems as well, to be presented in forthcoming publications.

Despite the fact that we are running our solver on digital computers (using a Runge-Kutta 
algorithm with adaptive step-size, 5th order Cash-Karp method for solving the ODEs) and
not on an analog device, it still shows superior performance on {\it very hard} MaxSAT 
problems, compared to competition solvers. This is because analog dynamical systems 
represent an entirely novel family of solvers and search dynamics, and for this reason 
they behave differently and thus can perform better than existing algorithms on 
hard problems or certain classes of hard problems. 
Trying to solve large, but simple MaxSAT problems on digital computers with this method, 
however, would show a weaker performance than digital MaxSAT solvers, simply 
because evolving a large number of coupled ODEs on digital machines is costly.
Direct hardware implementations, however, for example, via analog circuits are expected 
to run orders of magnitude faster than any state-of-the-art SAT solver run on the latest digital
computers, as shown in \cite{Xunzhao}. One reason for this is that in such analog 
circuits the von Neumann bottleneck 
is eliminated, with the circuit itself serving its own processor and memory, see 
\cite{Xunzhao} for details. It is also important to note that the system (\ref{sdyn1}-\ref{adyn1})
is not unique, other ODEs can be designed with similar or even better properties. This is
useful,  because the form given in (\ref{sdyn1}-\ref{adyn1}) is not readily amenable to 
simple hardware implementations, due to the constantly growing auxiliary variable dynamics 
(all variables represent a physical characteristic such as a voltage or a current and thus they
will have to have an upper limit value for a given device). However, the auxiliary variables
do not need to grow always exponentially, there are other solver variants in which
they only grow exponentially when needed, otherwise they can decay as well 
\cite{ModSystem}, allowing for better hardware implementations. 

To illustrate the effectiveness of our solver, as an application, we used it to find 
colorings in the famous two-color Ramsey problem and in particular for $R(5,5)$, 
which is still open.  We have shown, that the
two-color Ramsey problem avoiding monochromatic $m$-cliques can be translated into
an $\frac{m(m-1)}{2}$-SAT problem and thus into a $10$-SAT problem for $m=5$. 
Note that most digital SAT solving algorithms focus on 3-SAT or 4-SAT problems, and usually are
unable to handle directly the much harder $10$-SAT.   
Our solver when run on the corresponding $10$-SAT (or $10$-MaxSAT) was able 
to find colorings of the complete graph of order 42 avoiding monochromatic 5-cliques, 
and a coloring with only two (!) monochromatic 5-cliques on 6 nodes for the complete graph 
on 43 vertices (existing colorings in the literature for $N=43$ quote 500+ monochromatic 
5-cliques), adding further support to the conjecture that $R(5,5) = 43$. 

\section*{\large Methods} \label{Methods}

\subsection*{1. Algorithm description} 
Here we give a simple, non-optimized variant of the algorithm (see flowchart in Fig. S2).
Certainly, better implementations can be devised, for example with 
better fitting routines, however the description below is easier to follow
and works well.  Given a SAT problem, we first determine the $b$ 
parameter as described previously, then:
\begin{itemize}
\item[1.] Initially we set $\overline{E}=M$, $\Gamma_{min}$, 
$\Gamma_{max} \gg \Gamma_{min}$, $\Gamma^{pred}\left(\overline{E}-1\right)=
\Gamma_{min}+1$ and $t_{max}$.  Unless specified otherwise, in our simulations we used $\Gamma_{min}=100$,
$\Gamma_{max}=2\times 10^6$, $t_{max} = 50$. 
\item[2.]  To initialize our statistics, we run $\Gamma_{min}$ trajectories
up to $t_{max}$, each from a random initial condition. For every such 
trajectory $\omega$ we update the $p(E,\,t)$ distributions as function
of the energies of the orthants visited by $\omega$. We record the lowest 
energy value found $\overline{E}(\Gamma_{min})$.
\item[3.] Starting from $\Gamma = \Gamma_{min} + 1$ and up to $\Gamma_{max}$, we 
continue running trajectories in the same way and for each one of them check:
 \begin{itemize}
\item[a)] if $E_s \leq \overline{E}$, set $\overline{E}\defeq\min(E_s, \overline{E})$, update  $p(E,\,t)$ and go to step 4.
\item[b)] if $\Gamma$ just reached $\Gamma^{pred}\left(\overline{E}-1\right)$, go to step 4. 
\item[c)] if $\Gamma = \Gamma_{max}$, output ``Maximum number of steps reached, increase
$\Gamma_{max}$.'', output the lowest energy value found, the predicted $E^{pred}_{min}$ and the quality of fit for $E^{pred}_{min}$, then halt.
\end{itemize}
\item[4.] Using the $p(E,\,t)$ distributions, estimate the escape rates $\kappa(E)$ as described in Methods section 2.
\item[5.] The $\kappa(E)$ curve is extrapolated to the $E-1$ value obtaining $\kappa(E-1)$ 
and then using this we predict $\Gamma^{pred}\left(\overline{E}-1\right)$ (Methods section 3).   Further extrapolating the $\kappa(E)$ curve to $\kappa = 0$ we obtain $E^{pred}_{min}$ (see Methods section 4).
\item[6.] We check the consistency of the prediction defined here as saturation of the predicted values. We call it consistent, if $E^{pred}_{min}$ has not changed during the last 5 predictions. If it is not consistent yet, we go to step 4 and continue running new trajectories. If the prediction is consistent, we check for the following halting conditions:
    \begin{itemize}
    \item if $E^{pred}_{min}= \overline{E}(\Gamma)$  then we decide the global optimum has been found: $E^{dec}_{min}= E^{pred}_{min}= \overline{E}(\Gamma)$ and go to step 7.
    \item if the fitting is consistently predicting $E^{pred}_{min}> \overline{E}(\Gamma)$ (usually it is very close, $\overline{E}(\Gamma)+1$) we check the number of trajectories that
has attained states with $\overline{E}(\Gamma)$, i.e., $n(\overline{E})=\left[1-p(\overline{E}(\Gamma),\,t_{max})\right] \Gamma$. If it is large enough (e.g. $> 100$), we decide to stop running new trajectories and set $E^{dec}_{min}= \overline{E}(\Gamma)$ and go to step 7.
    \item if $E^{pred}_{min}<\overline{E}(\Gamma)$  then we most probably have not  found the global optimum yet and we go to step 4.
    \end{itemize}
We added additional stopping conditions that can shorten the algorithm in case of easy problems, see Methods section 5, but these are not so relevant.
\item[7.] The algorithm ends and outputs $E^{pred}_{min}$, $E^{dec}_{min}$, $\overline{E}$ values, the Boolean variables corresponding to the optimal state found, along with the quality of fit.
\end{itemize}

\subsection*{2. Estimation of the escape rates $\kappa(E)$} 

As seen in  Fig. \ref{fig2} and Fig. S1 the exponential decay of the $p(E,\,t)$ distribution settles in after a short transient period. Theoretically the escape rate can be obtained by fitting the exponential on that last part of the curves (Fig. \ref{fig2}a). However, while running the algorithm it would be difficult to automatically estimate the region where the exponential should be fitted. The simple approach that works well is to estimate the escape rates as $\kappa(E)\simeq -\ln(p(E,\,t_{max})/t_{max}$, which practically would correspond to the exponential behavior being valid on the whole $(0,t_{max})$ interval. Note, the $p(E,\,t)$ is a cumulative distribution with $p(E,\,0) = 1$. Usually this estimation is very close to the fitted values (Fig. \ref{fig2}b), but notice that what matters here is the scaling behavior of the escape rates, and this is quite precisely obtained this way because it simply uses the scaling behavior of the $p(E,\,t_{max})$ values, instead of the fittings, which is sensitive to the chosen interval.

 \subsection*{3. Predicting the number of trajectories needed to find a lower energy}
 After calculating the escape rates $\kappa(E)$ one can estimate the number of expected trajectories needed to find a lower energy value: $\Gamma^{pred}(\overline{E}-1)$ as described below. Clearly, $p(E,\,t_{max})$ is the probability that a trajectory has not reached the energy level $E$ up to time $t_{max}$.  This means that $1-p(E,\,t_{max})$ is the probability that a trajectory did reach energy $E$, up to time $t_{max}$. Running $\Gamma$ trajectories, thus $n(\overline{E})=[1-p(E,t_{max})] \Gamma$ will give the expected number of trajectories that reached energy $E$. Thus, the expected number of trajectories we need to run {\it in total} to find the $\overline{E}-1$ energy value at least once is:
 \begin{equation}
 \Gamma^{pred}(\overline{E}-1)=\frac{1}{1-p(t_{max},\;\overline{E}-1)}\;.
 \end{equation}  
 However, no trajectory has reached energy $\overline{E}-1$ yet, and thus we don't have
$p(\overline{E}-1,\,t_{max})$. Instead, it is computed from 
$p(\overline{E}-1,\,t_{max}) \simeq e^{-\kappa(\overline{E}-1)t_{max}}$, after extrapolating
the $\kappa(E)$ curve to obtain $\kappa(\overline{E}-1)$.

\subsection*{4. Predicting the global optimum}

When fitting the curve $E=E_0+a\kappa^{\beta}$ on our data points we used the Numerical Recipes implementation \cite{NumRec92} of the Levenberg-Marquardt non-linear curve fitting method \cite{Levenberg44,Marquardt63}. This implementation has some weaknesses, one could choose to use other implementations or other methods. Sometimes a 3-parameter  fitting is too sensitive and does not give good results. Because we do not need a very precise value for $E_0$ (we just need to find the integer interval it falls into, because $E_{min}^{pred}=[E_0]+1$, $[.]$ meaning the integer part) we perform a series of fittings always  fixing $E_0$ and leaving only 2 unknown parameters ($a,\beta$). For each $E_0=\overline{E}, \overline{E}-0.1, \overline{E}-0.2,....$ we then perform a fitting and check the $\chi^2$ error.  The fitting with minimal error is chosen as the final $E_0$ and final fitted curve.

\subsection*{5. Additional stopping conditions }

 There are cases usually for easy problems, when the fitting using the form \eqref{plfit} does not work well, but based on certain simple conditions we can trust that the global optimum has been found. For example: 
 \begin{itemize}
 \item  if there are not enough (e.g less than 5) data points in the $\kappa(E)$ curve fitted (this partly explains why the fitting does not give good prediction), but the lowest energy has already been found many times ($n(\overline{E})>n_{max}$, e.g. $n_{max}=1000$). This happens for very easy problems. 
\item The fitting is consistently predicting another  $E^{pred}_{min}\neq \overline{E}$, but  $n(\overline{E})$ is very large and $\Gamma>\Gamma^{pred}(\overline{E}-1)$, so according to the dynamics, it seems a lower energy should have been found already.
\end{itemize}
In such cases we exit the algorithm (step 7) with the decision:  $\overline{E}\neq E^{pred}_{min}, E^{dec}_{min}=\overline{E}$. 


\section*{\large Acknowledgments}
We thank to Xiaobo S. Hu, S. Datta, S. Basu,
Z. N\'eda, E. Regan, X. Yin,  F. Moln\'ar and S. Kharel 
for useful comments and discussions. We also thank Brendan D. MacKay for pointing out 
errors in some of the SI tables in the first version of the article. 
This work was supported in part by a grant 
of the Romanian National Authority for Scientific Research and Innovation 
CNCS/CCCDI-UEFISCDI, project nr. PN-III-P2-2.1-BG-2016-0252 (MER), the 
GSCE-30260-2015 Grant for Supporting Excellent Research of the Babe\c{s}-Bolyai 
University (BM, MER). It was also supported in part by the National Science
Foundation under Grants CCF-1644368 and 1640081, and by the Nanoelectronics
Research Corporation, a wholly-owned subsidiary of the Semiconductor
Research Corporation, through Extremely Energy Efficient Collective
Electronics, an SRC-NRI Nanoelectronics Research Initiative under Research
Task ID 2698.004 (ZT, MV).

\section*{\large Author Contributions} 
The authors designed the methods and algorithm together, contributed analysis and tools equally. 
Both ZT and MER wrote and reviewed the manuscript.

\section*{\large Additional Information} 

{\bf Competing financial interests} 
The authors declare no competing financial interests. 


\renewcommand{\thefootnote}{\roman{footnote}}
\renewcommand{\thefigure}{S\arabic{figure}}
\setcounter{figure}{0}

\section*{\Large \sc \centerline{Supplementary Information}}

\vfill

\subsection*{\large \underline{Supplementary Figures}}

\vspace*{1cm}

\begin{figure}[!htb] 
\centering \includegraphics[width=1.00\textwidth]{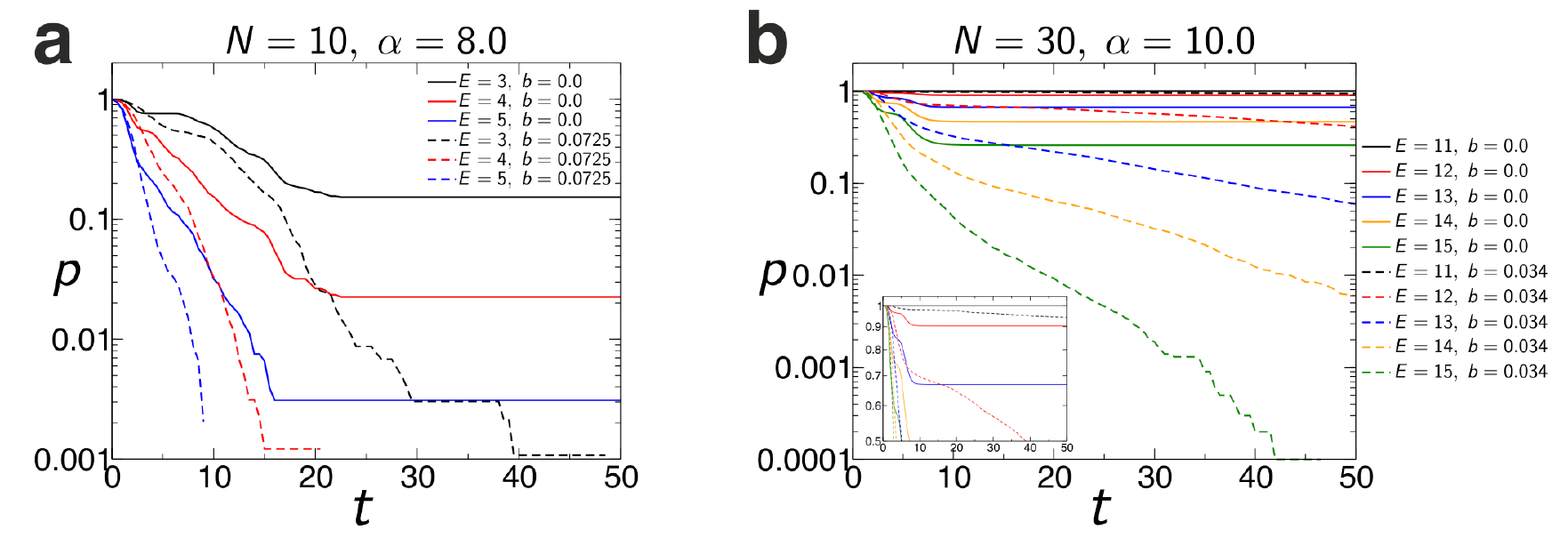}
\caption{{\bf The $p(E,t)$ distribution of transient times compared between the SAT-solver algorithm presented in \cite{NatPhys_ET11} and the new max-SAT solver dynamics.} $p(E,t)$ is the probability that up to time $t$ a trajectory has not yet visited an orthant with energy smaller than E. As shown for two max-SAT instances (a) $N=10$, $\alpha = 8.0$, $b=0.0725$, $E_{min} = 3$. (b) $N=30$, $\alpha = 10.0$, $b=0.034$, $E_{min} = 11$, when $b=0$ (the original analog SAT solver dynamics) the exponential decay of the distribution stops at relatively small $t$ indicating that many trajectories have been trapped at $\bm{s}=0$. When $b > 0$  the distribution is much steeper and the trapping phenomenon does not occur.
} \label{figS1} 
\end{figure}

\vfill
\vfill

\clearpage
\newpage

\begin{figure}[!htb] 
\centering \includegraphics[width=0.90\textwidth]{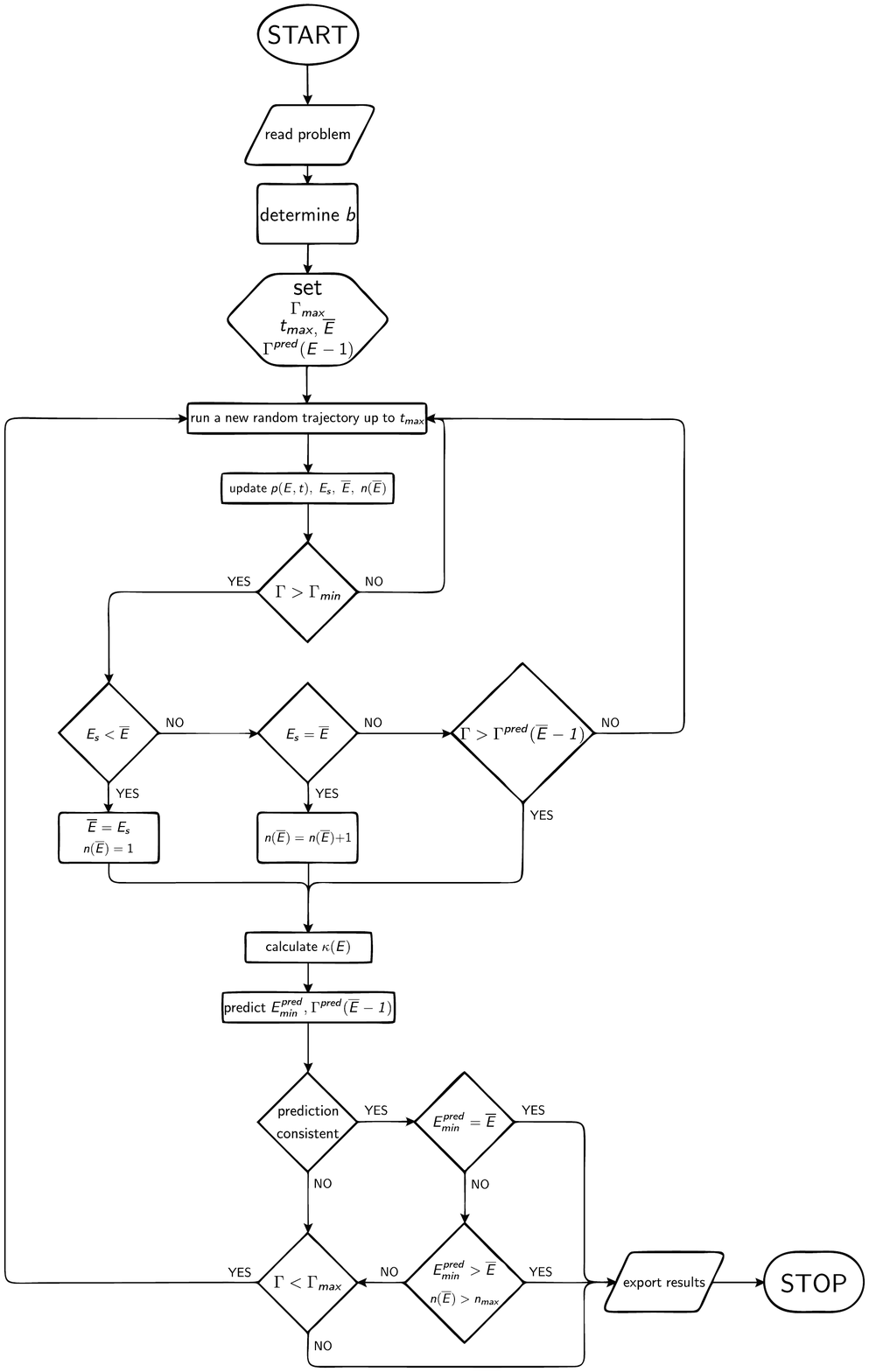}
\caption{{\bf Flowchart of the dynamics described in detail in the Methods section.} } \label{figS2} 
\end{figure}

\clearpage
\newpage

\begin{figure}[!htb] 
\centering \includegraphics[width=1.0\textwidth]{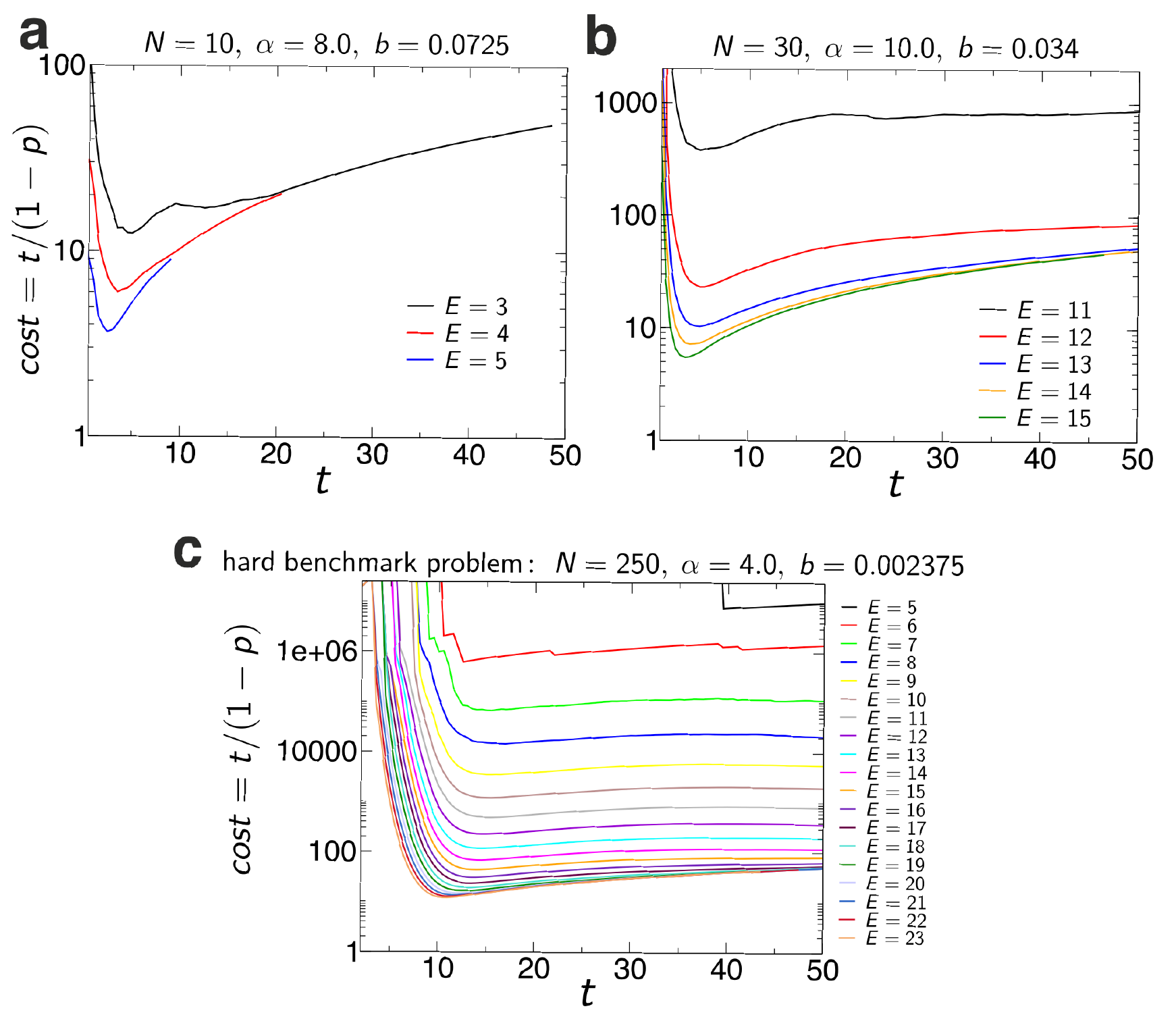}
\caption{{\bf The cost of finding an energy level as function of time.} If we run $\Gamma$ trajectories, each one up to time $t$, the total cost is proportional with $\Gamma t$. The number of times an energy level was found is equal to $\left(1-p(E,t) \right)\Gamma$, so we estimate the cost of finding a state with energy $E$  as $t/\left(1-p(E,t) \right)$. We plot this cost function for different energy levels for three different MaxSAT instances. (a) $N=10$, $\alpha = 8.0$, $b=0.0725$, $E_{min} = 3$. (b) $N=30$, $\alpha = 10.0$, $b=0.034$, $E_{min} = 11$. (c) The hard benchmark problem presented in detail in Figs. \ref{fig2}, \ref{fig4} of the main text with $N=250$, $\alpha = 4.0$, $b=0.002375$, $E_{min} = 5$. The curves show a minimum at a relatively low $t$, indicating that it is much more efficient to run many short trajectories, than running a few for longer times. In case of random SAT instances with $N \leq 100$ we usually choose $t_{max} = 25$. For larger instances such as the benchmark problem we used $t_{max} = 50$ in our algorithm.} \label{figS3} 
\end{figure}

\clearpage
\newpage

\begin{figure}[!htb] 
\centering \includegraphics[width=1.00\textwidth]{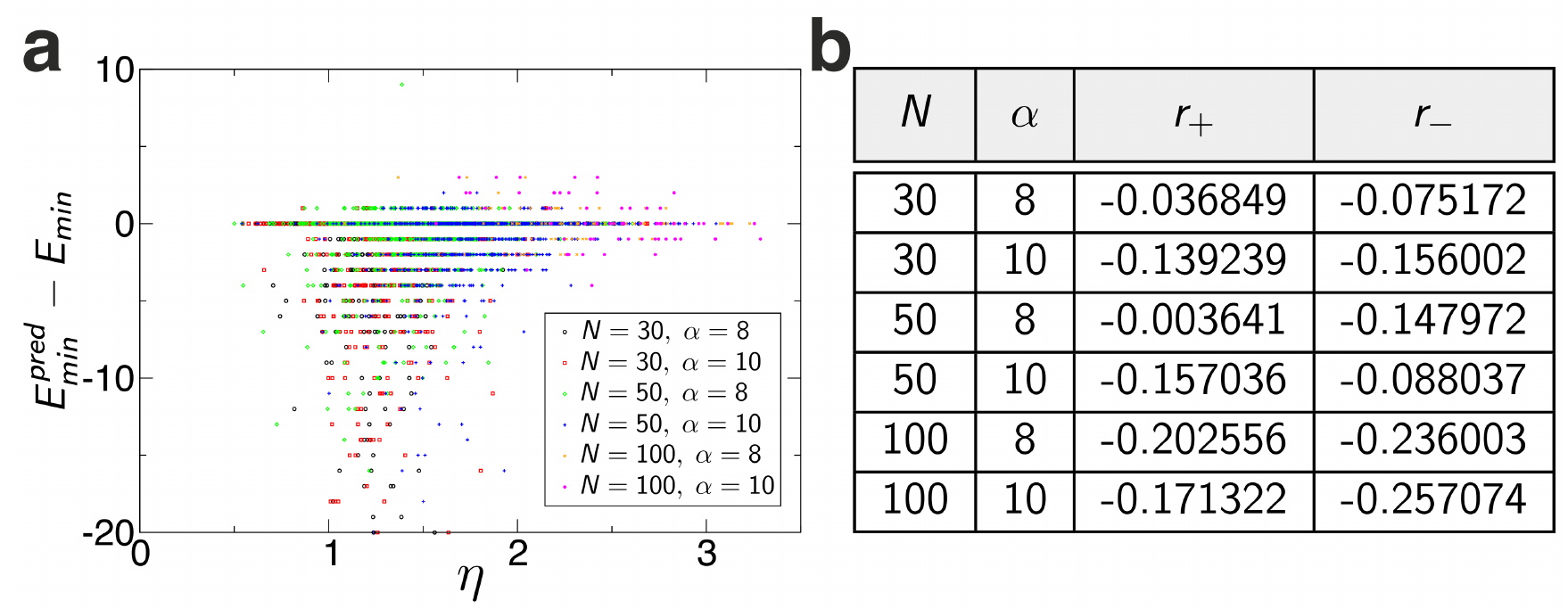}
\caption{{\bf Correlation between the error of prediction and hardness of random 3-MaxSAT instances.} (a) The difference between the predicted and real global minimum value as function of the hardness measure $\eta = -\log \kappa / \log N$, which is applicable for individual instances. We use different symbols and colors for instances with different $N$ and $\alpha$ (see legend). Large errors occur mainly in easy problems, at small $\eta$ values. (b) This is also shown by the negative values of the Pearson correlation coefficients obtained between the absolute value of errors and $\eta$, calculated separately for positive $(r_+)$ and negative errors $(r_-)$. } \label{figS4} 
\end{figure}

\clearpage
\newpage

\begin{figure}[!htb] 
\centering \includegraphics[width=1.00\textwidth]{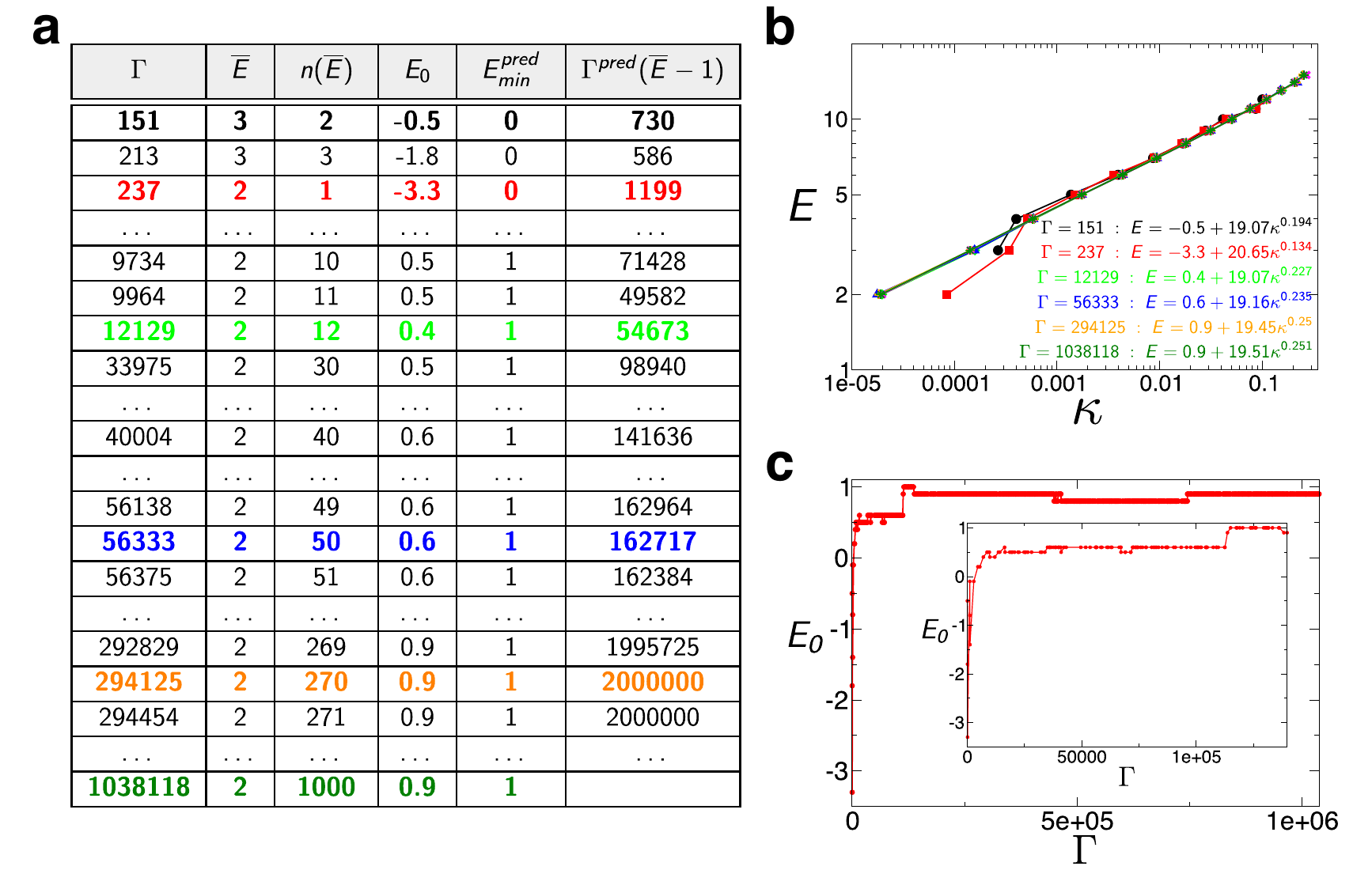}
\caption{{\bf Performance of our algorithm on the 4-SAT problem {\tt HG-4SAT-V150-C1350} {\tt-24.cnf} listed in the table of Fig. 5 of the main text.} Similarly, to Fig. \ref{fig4} of the main text in (a) we show the values of the relevant measures at each fitting performed. The fitting for each colored line is shown in (b). (c) The parameter $E_0$ that provides the final prediction of the global optimum is shown as function of $\Gamma$, the number of trajectories ran. In this case our predicition slightly underestimates the minimum, predicting 1. However, after the energy state 2 is found 1000  times we stop the search.} \label{figS5} 
\end{figure}

\clearpage
\newpage

\begin{figure}[!htb] 
\centering \includegraphics[width=1.00\textwidth]{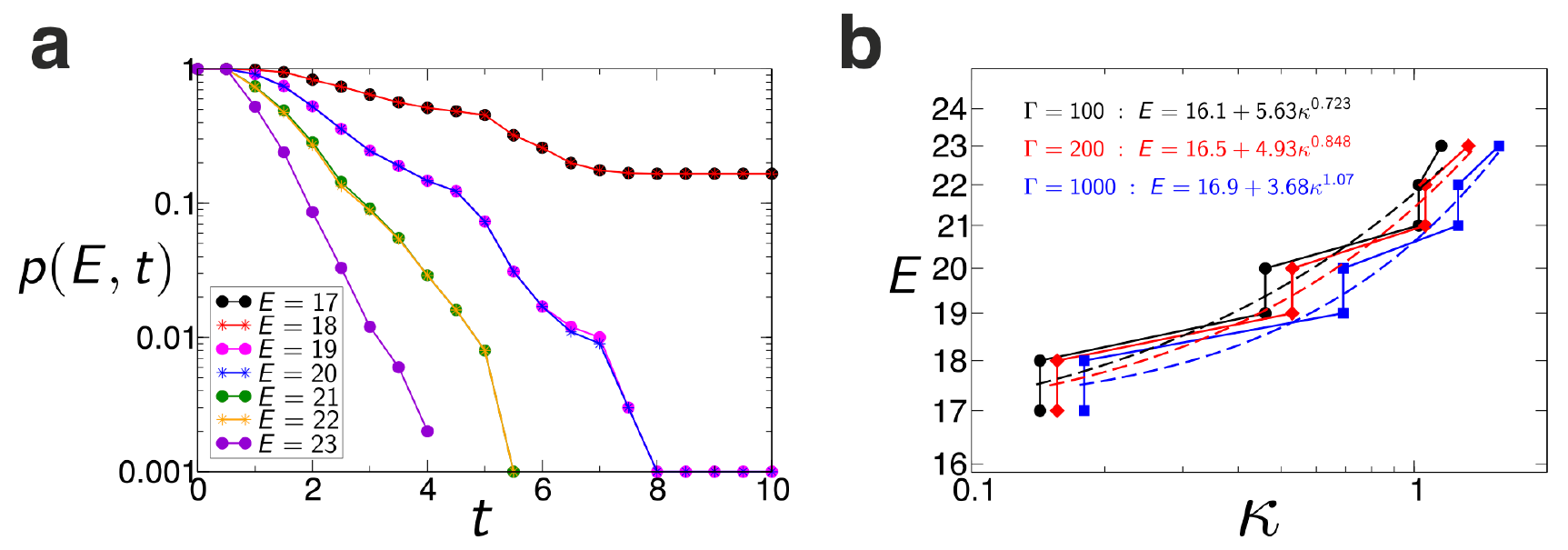}
\caption{{\bf Predicting the global optimum in a spin-glass hard benchmark problem.}  $k=2$, $N=27$, $M=162$, $E_{min} = 17$, $b=0.042438$ (the problem instance {\tt t3pm3-5555.spn.cnf} can be downloaded from \cite{benchmark14}). Note, Max-2-SAT is also NP-hard, unlike 2-SAT, which is in P. (a) The $p(E,t)$ distributions after running $\Gamma = 1000$ trajectories. Because of the special structure of the spin-glass problem pairs of consecutive energy levels show the same distributions (a single spin-flip can change the energy only in units of 2). (b) The energy $E$ as function of the estimated $\kappa(E)$ values after running $\Gamma = 100,200,1000$ trajectories (see legend). The fitted curves are shown with dashed lines. $E_0 \in [16,17)$ indicating a correct prediction already after $\Gamma = 100$ trajectories. This is an easy problem for our algorithm and $E_{min} = 17$ is found 79 times by the first $\Gamma = 100$ trajectories.
} \label{figS6} 
\end{figure}

\subsection*{\large \underline{Supplementary tables and data}}

\renewcommand{\thefootnote}{\roman{footnote}}
\renewcommand{\thetable}{S\arabic{table}}
\setcounter{table}{0}

\begin{table}[h] 
\caption{{\bf The 3-SAT problem shown in Figure 1 of the main text.} We have  $N=10$ variables and  $M=80$ clauses (constraints) in conjunctive normal form. Each clause is shown in parenthesis as the series of 3 variables or their negation (indicated by negative sign).  For example, the first triple in parantheses indicates the clause/constraint: ($x_5$ OR (NOT $x_8$) OR $x_9$).} \label{tableS1} \sf
\begin{tabular}{l} 
(5,-8,9)(-1,-3,-7)(9,4,-8)(-1,-9,4)(7,2,3)(9,5,4)(8,9,-3)(10,-5,9)(9,7,8)(3,1,6) (7,10,3)\\
(-5,10,-3)(-7,6,4)(-8,1,-10)(-1,-2,3)(-9,-2,-3)(5,7,8)(-5,-3,4)(9,-2,1)(-3,-1,-7)(10,5,4)\\
(-7,-10,-4)(-9,-10,3)(2,-1,10)(-5,-10,-7)(-9,6,8)(-9,-4,-8)(-5,-3,-8)(-9,3,-7)(-6,2,5)\\
(-2,1,-8)(1,6,9)(5,-9,2)(10,-1,7)(5,-1,-3)(6,-7,2)(8,-5,7)(-8,-7,-3)(4,-7,3)(4,-9,2)(1,6,-7)\\
(-9,-2,5)(10,-4,-5)(4,-2,-9)(-7,2,1)(4,2,-8)(-2,-10,-5)(6,-3,7)(-1,-3,7)(-1,6,4)(-9,-4,3)\\
(-4,10,-5)(9,6,-2)(-8,-2,5)(2,-1,3)(-6,-4,10)(7,-5,2)(7,3,-5)(-7,9,-6)(-4,6,2)(-6,9,-5)\\
(-10,-1,2)(5,-8,-7)(8,7,-2)(-8,-2,1)(6,1,-8)(8,5,-2)(-8,3,6)(10,2,-3)(9,-7,2)(-6,10,2)\\
(1,-3,4)(6,2,-8)(9,2,10)(2,5,-1)(-1,8,4)(-3,1,-4)(-10,9,-7)(-4,-5,-9)(-6,-7,10)
\end{tabular}
\end{table}

\noindent The problem instances with the solutions presented in Tables 
\ref{tableS2}-\ref{tableS13} below can all be downloaded from \cite{benchmark14}, 
\cite{benchmark16}.


\begin{table}[h] 
\caption{{\bf The optimal solution (list of Boolean variables corresponding to the optimal state found) 
found for the {\tt HG-3SAT-V250-C1000-1.cnf} MaxSAT competition problem: $E_{min} = 5$. }} 
\label{tableS2} \sf
\begin{tabular}{l} 
0 1 1 1 1 0 1 1 0 1 1 1 1 1 0 0 0 1 0 0 0 0 0 1 1 1 1 1 1 1 0 1 1 0 0 0 0 1 1 0 0 1 0 0 0 0 0 1 1 1 \\
0 0 1 1 1 0 1 1 1 1 1 0 0 1 1 0 0 0 0 0 1 0 0 0 0 1 1 0 1 0 1 0 0 0 1 1 1 0 0 1 0 0 0 1 1 1 1 0 1 0 \\
0 0 1 0 0 0 0 0 0 0 0 0 0 0 0 1 1 0 0 1 0 0 1 1 0 1 0 0 0 0 1 1 1 0 0 0 1 0 0 1 0 1 1 0 0 0 1 1 1 0 \\
0 0 1 0 0 1 1 1 0 1 1 0 1 1 1 0 0 0 1 0 0 0 0 1 0 1 1 1 0 1 1 1 1 1 0 1 0 1 0 0 0 0 1 0 0 1 1 0 0 1 \\
0 0 1 1 1 0 0 1 1 1 1 1 0 0 1 1 0 0 1 0 1 0 1 0 1 0 0 0 0 0 1 0 1 0 1 1 0 0 1 0 0 0 1 0 1 0 1 0 0 1
\end{tabular}
\end{table}

\begin{table}[h] 
\caption{{\bf {\tt HG-3SAT-V250-M1000-100.cnf} optimal solution found: $E_{min} = 7$. \hfill} } 
\label{tableS3} \sf
\begin{tabular}{l} 
1 1 1 0 1 1 1 0 0 1 0 1 0 1 1 1 1 0 0 0 1 0 0 1 0 0 1 1 1 1 1 0 1 1 1 1 0 0 1 1 1 1 1 1 0 0 1 0 1 0 \\
0 0 1 0 0 1 0 1 1 1 0 1 0 0 1 1 1 1 0 1 1 0 0 0 1 1 1 1 1 1 1 0 0 0 1 0 1 1 1 1 1 0 1 1 0 0 0 0 1 0 \\
1 0 1 1 0 0 0 1 1 0 1 0 0 0 1 0 1 0 0 0 1 1 1 1 0 0 0 0 0 0 1 0 1 0 1 0 0 0 1 0 1 0 0 0 1 0 0 1 0 1 \\
1 0 0 1 0 1 1 0 0 1 0 1 0 1 1 1 1 1 1 1 1 0 0 0 0 1 0 0 1 0 1 1 1 0 0 0 0 1 0 1 1 0 0 0 1 1 0 1 1 0 \\
0 0 1 1 0 0 1 0 1 0 0 0 1 1 1 1 1 0 0 0 0 1 1 1 1 0 1 0 0 0 0 1 1 0 0 0 1 1 1 0 0 0 0 0 0 1 1 0 1 1
\end{tabular}
\end{table}

\begin{table}[h] 
\caption{{\bf {\tt HG-3SAT-V300-M1200-9.cnf} optimal solution found: $E_{min} = 8$. \hfill}} 
\label{tableS4} \sf
\begin{tabular}{l} 
1 1 1 1 1 0 1 1 0 0 1 0 0 1 0 0 1 1 0 1 1 1 1 0 1 0 0 0 0 0 1 0 0 0 0 0 1 1 1 1 0 1 1 0 1 1 1 0 1 0 0 \\
0 1 1 0 1 1 0 0 1 0 1 0 1 1 1 0 0 1 1 0 0 1 0 1 0 0 0 1 0 1 0 1 1 0 0 0 1 0 0 1 0 0 1 1 1 1 1 0 1 1 0 \\
0 1 1 1 1 1 1 1 1 1 0 0 1 0 1 0 0 0 0 0 0 1 0 0 1 1 1 0 1 0 0 1 1 0 0 0 0 0 1 1 0 1 0 0 1 1 1 1 0 1 1 \\
0 1 1 1 0 0 1 0 1 0 0 0 0 0 1 0 0 0 1 1 0 1 1 1 0 0 0 1 1 1 1 1 0 0 0 0 0 0 0 0 0 1 0 1 1 0 0 0 1 0 1 \\
0 1 1 1 0 1 1 0 1 1 0 1 0 1 1 1 1 0 1 1 1 0 0 0 1 1 0 1 0 1 1 0 1 0 0 1 1 0 1 0 1 1 0 0 0 0 0 1 1 0 1 \\
1 1 0 0 1 0 0 1 0 0 0 1 1 1 1 0 0 0 0 0 0 0 1 0 1 0 1 0 1 1 0 0 0 1 0 1 1 1 1 0 0 1 1 1 1
\end{tabular}
\end{table}

\begin{table}[h] 
\caption{{\bf {\tt HG-4SAT-V100-M900-2.cnf} optimal solution found:  $E_{min} = 2$. \hfill}} 
\label{tableS5} \sf
\begin{tabular}{l} 
1 1 0 1 1 1 0 1 1 1 1 1 1 0 1 0 1 0 0 0 0 0 1 1 1 1 0 0 1 0 0 1 0 0 1 1 1 0 1 1 0 0 0 0 0 1 1 1 0 0 1 \\
0 1 1 0 1 1 0 1 1 1 0 0 0 0 0 0 1 1 1 0 0 0 1 0 0 1 0 1 0 1 0 1 1 1 0 0 0 0 1 1 1 0 1 1 0 0 1 0 1
\end{tabular}
\end{table}

\begin{table}[h] 
\caption{{\bf {\tt HG-4SAT-V100-M900-4.cnf} optimal solution found:  $E_{min} = 2$. \hfill}} 
\label{tableS6} \sf
\begin{tabular}{l} 
0 0 1 1 1 1 0 0 1 1 0 0 0 0 0 0 0 1 0 0 1 1 1 0 1 1 0 1 1 0 0 1 0 1 1 1 0 0 1 1 0 1 0 1 0 1 1 0 1 1 1 \\
0 0 1 0 0 1 0 1 0 0 0 0 0 0 1 0 0 1 1 0 1 0 0 0 1 1 1 0 1 0 1 0 0 1 0 0 1 1 1 0 0 0 0 1 1 0 0 0 0
\end{tabular}
\end{table}

\begin{table}[h] 
\caption{{\bf  {\tt HG-4SAT-V100-M900-7.cnf} optimal solution found:  $E_{min} = 2$. \hfill}} 
\label{tableS7} \sf
\begin{tabular}{l} 
0 1 1 1 1 0 0 0 1 0 1 0 1 0 0 0 0 1 1 1 0 1 0 1 0 1 1 1 0 0 0 0 1 0 1 0 1 1 0 0 0 0 1 0 1 0 0 1 1 0 1 \\
1 1 1 0 1 1 0 0 1 1 1 0 1 1 0 1 0 1 1 0 0 1 1 0 1 1 1 1 0 1 0 1 1 0 1 1 0 0 1 1 1 1 1 1 1 0 0 0 1 
\end{tabular}
\end{table}

\begin{table}[h] 
\caption{{\bf  {\tt HG-4SAT-V100-M900-14.cnf} optimal solution found: $E_{min} = 2$. \hfill}} 
\label{tableS8} \sf
\begin{tabular}{l} 
0 0 1 0 0 0 1 1 1 1 1 1 1 1 0 0 1 0 0 0 1 0 1 1 0 0 0 1 1 1 0 1 1 0 1 0 0 1 0 0 1 0 0 1 1 1 1 0 0 0 1 \\
1 0 0 1 0 1 0 0 0 0 0 0 0 0 1 1 0 0 0 0 1 1 1 0 1 1 1 0 1 1 0 0 0 0 1 1 0 0 0 0 1 1 1 0 0 1 1 1 1
\end{tabular}
\end{table}

\clearpage
\newpage

\begin{table}[h] 
\caption{{\bf  {\tt HG-4SAT-V100-M900-19.cnf optimal solution found:}  $E_{min} = 2$. \hfill}} 
\label{tableS9} \sf
\begin{tabular}{l} 
0 1 1 1 0 1 0 0 1 1 1 0 0 0 0 0 1 1 1 1 1 0 1 0 1 0 0 1 0 0 0 1 0 0 0 0 0 0 1 0 1 1 1 1 1 0 1 1 1 0 1 \\
1 1 1 1 1 1 1 1 1 1 0 0 0 0 1 0 1 1 1 0 0 1 0 1 1 1 1 1 0 0 0 0 0 1 0 1 1 0 0 0 1 0 1 1 1 0 0 1 1
\end{tabular}
\end{table}

\begin{table}[h] 
\caption{{\bf {\tt HG-4SAT-V100-M900-20.cnf optimal solution found:} $E_{min} = 2$. \hfill}} 
\label{tableS10} \sf
\begin{tabular}{l} 
1 0 0 1 0 1 0 0 0 0 0 0 0 0 1 0 1 1 0 0 1 1 1 1 0 1 0 1 1 1 0 1 1 0 1 0 1 1 0 0 1 1 0 0 0 0 1 1 1 0 0 1 \\
1 0 1 1 1 1 0 1 1 1 1 1 1 1 1 0 0 0 1 1 0 1 0 1 0 0 0 0 1 1 1 0 0 0 1 0 0 1 1 1 1 1 0 0 0 1 0 0
\end{tabular}
\end{table}

\begin{table}[h] 
\caption{{\bf {\tt HG-4SAT-V100-M900-23.cnf} optimal solution found: $E_{min} = 2$. \hfill}} 
\label{tableS11} \sf
\begin{tabular}{l} 
0 0 0 0 1 0 1 0 1 0 0 0 0 1 0 1 1 0 0 0 0 1 0 1 1 0 1 0 1 0 1 0 1 0 1 1 0 1 0 0 0 1 1 0 0 1 1 0 0 0 1 1 \\
1 0 1 0 1 0 1 1 1 0 1 0 1 0 1 0 1 0 1 1 0 0 0 1 0 0 1 1 1 0 1 1 1 1 0 0 0 0 0 0 1 0 1 0 0 0 0 1
\end{tabular}
\end{table}

\begin{table}[h] 
\caption{{\bf {\tt HG-4SAT-V150-M1350-23.cnf} optimal solution found: $E_{min} = 0$. \hfill}} 
\label{tableS12} \sf
\begin{tabular}{l} 
0 1 1 0 0 0 0 1 1 0 1 1 1 0 0 0 0 1 1 0 1 1 0 1 1 1 0 0 1 1 1 0 1 0 1 0 0 1 0 0 0 0 0 0 1 0 0 1 0 1 0 1 1 \\
0 1 0 0 0 1 0 1 0 1 1 0 0 1 1 0 0 1 0 0 0 0 0 1 1 1 0 1 0 1 0 1 1 0 1 1 1 1 0 1 1 0 0 1 0 1 1 1 1 1 0 0 0 \\
1 1 0 1 0 1 1 1 1 1 1 0 1 1 1 0 0 0 0 0 0 1 0 1 0 1 0 1 1 0 1 0 0 0 0 1 1 1 1 0 0 1 1 0
\end{tabular}
\end{table}

\begin{table}[h] 
\caption{{\bf {\tt HG-4SAT-V150-M1350-24.cnf} optimal solution found:  $E_{min} = 2$. \hfill}} 
\label{tableS13} \sf
\begin{tabular}{l} 
0 0 1 0 1 0 1 0 0 0 0 1 1 1 0 0 0 1 0 1 0 0 1 1 1 0 0 0 0 1 1 0 1 1 1 1 0 1 1 0 1 0 0 0 1 0 0 0 0 1 0 0 \\
0 0 0 0 1 0 0 0 0 1 0 0 0 0 1 0 1 1 1 1 1 0 1 1 0 1 0 0 1 0 1 0 0 0 1 1 0 0 1 0 1 0 1 0 1 1 1 1 1 1 1 1 \\
1 0 1 0 0 1 1 1 1 1 1 0 0 1 0 1 1 1 0 1 0 1 0 0 0 0 0 1 1 1 0 1 1 0 1 1 0 0 0 1 0 0 0 1 0 0
\end{tabular}
\end{table}

\begin{table}[htbp] 
\caption{{\bf Spin-glass problem {\tt t3pm3-5555.spn} presented in Fig \ref{figS6}. Optimal solution found: $E_{min} = 17$. \hfill}} \label{tableS14} \sf
\begin{tabular}{l} 
0 0 0 1 0 1 1 0 1 1 0 1 1 1 0 0 1 0 0 1 1 1 1 1 0 0 0 \qquad  \qquad \qquad \qquad \qquad  \qquad \qquad \\
 \qquad  \qquad \qquad \qquad \qquad \qquad \\
\end{tabular}
\end{table}

\vfill
\vfill

\clearpage
\newpage

\begin{table}[t!]
\caption{{\bf The matrix shown in Fig. \ref{fig7}b of the main text that gives a complete Ramsey coloring of a complete graph on 42 nodes. There are no monochromatic 5-cliques ($E_{min} = 0$).}} 
\label{tableS15} \begin{singlespacing}  \sf
\begin{tabular}{l} 
0 0 1 1 1 0 0 0 0 0 0 0 1 1 1 0 1 1 0 1 1 0 1 1 1 0 1 1 0 1 1 1 0 0 0 0 0 0 0 1 1 0 \\
0 0 0 1 1 1 0 0 0 0 0 0 0 1 1 1 0 1 1 0 1 1 1 1 1 1 0 1 1 0 1 1 1 0 0 0 0 0 0 0 1 1 \\
1 0 0 0 1 1 1 0 0 0 0 0 0 0 1 1 1 0 1 1 0 1 1 1 0 1 1 0 1 1 0 1 1 1 0 0 0 0 0 0 0 1 \\
1 1 0 0 0 1 1 1 0 0 0 0 0 0 0 1 1 1 0 1 1 0 1 1 0 0 1 1 0 1 1 0 1 1 1 0 0 0 1 0 0 0 \\
1 1 1 0 0 0 1 1 1 0 0 0 0 0 0 0 1 1 1 0 1 1 0 1 1 0 0 1 1 0 1 1 0 1 1 1 0 0 0 0 0 0 \\
0 1 1 1 0 0 0 1 1 1 0 0 0 0 0 0 0 1 1 1 0 1 1 0 1 1 1 1 1 1 0 1 1 0 1 1 1 0 0 0 0 0 \\
0 0 1 1 1 0 0 0 1 1 1 0 0 0 0 0 0 0 1 1 1 0 1 1 0 1 1 1 1 1 1 0 1 1 0 1 1 1 0 0 0 0 \\
0 0 0 1 1 1 0 0 0 1 1 1 0 0 0 0 0 0 0 1 1 1 0 1 1 0 1 1 0 0 1 1 0 1 1 0 1 1 1 0 0 0 \\
0 0 0 0 1 1 1 0 0 0 1 1 1 0 0 0 0 0 0 0 1 1 1 0 1 1 0 1 1 0 0 1 1 0 1 1 0 1 1 1 0 0 \\
0 0 0 0 0 1 1 1 0 0 0 1 1 1 0 0 0 0 0 0 0 1 1 1 0 1 1 0 1 1 0 1 0 1 0 1 1 0 1 1 1 0 \\
0 0 0 0 0 0 1 1 1 0 0 0 1 1 1 0 0 0 0 0 0 0 1 1 1 0 1 1 0 1 1 0 1 1 1 0 1 1 0 1 1 1 \\
0 0 0 0 0 0 0 1 1 1 0 0 0 1 1 1 0 0 0 0 0 0 0 1 1 1 0 1 1 0 1 1 0 0 1 1 0 1 1 0 1 1 \\
1 0 0 0 0 0 0 0 1 1 1 0 0 0 1 1 1 0 0 0 0 0 0 0 1 1 1 0 1 1 0 1 1 0 0 1 1 0 1 1 0 1 \\
1 1 0 0 0 0 0 0 0 1 1 1 0 0 0 1 1 1 0 0 0 0 0 0 0 1 1 1 0 1 1 0 1 1 0 1 1 1 0 1 1 0 \\
1 1 1 0 0 0 0 0 0 0 1 1 1 0 0 0 1 1 1 0 0 0 0 0 0 0 1 1 1 0 1 1 0 1 1 1 1 1 1 0 1 1 \\
0 1 1 1 0 0 0 0 0 0 0 1 1 1 0 0 0 1 1 1 0 0 0 0 0 0 1 1 1 1 0 1 1 0 1 1 1 0 1 1 0 1 \\
1 0 1 1 1 0 0 0 0 0 0 0 1 1 1 0 0 0 1 1 1 0 0 0 0 0 0 0 1 1 1 0 1 1 0 1 1 0 0 1 1 0 \\
1 1 0 1 1 1 0 0 0 0 0 0 0 1 1 1 0 0 0 1 1 1 0 0 0 0 0 0 0 1 1 1 0 1 1 0 1 1 0 0 1 1 \\
0 1 1 0 1 1 1 0 0 0 0 0 0 0 1 1 1 0 0 0 1 1 1 0 0 0 0 0 0 0 1 1 1 0 1 1 0 1 1 1 1 1 \\
1 0 1 1 0 1 1 1 0 0 0 0 0 0 0 1 1 1 0 0 0 1 1 1 0 0 0 0 0 0 0 1 1 1 0 1 1 0 1 1 1 1 \\
1 1 0 1 1 0 1 1 1 0 0 0 0 0 0 0 1 1 1 0 0 0 1 1 1 0 0 0 0 0 0 0 1 1 1 0 1 1 0 1 1 0 \\
0 1 1 0 1 1 0 1 1 1 0 0 0 0 0 0 0 1 1 1 0 0 0 1 1 1 0 0 0 0 0 0 0 1 1 1 0 1 1 0 1 1 \\
1 1 1 1 0 1 1 0 1 1 1 0 0 0 0 0 0 0 1 1 1 0 0 0 1 1 1 0 0 0 0 0 0 0 1 1 1 0 1 1 0 1 \\
1 1 1 1 1 0 1 1 0 1 1 1 0 0 0 0 0 0 0 1 1 1 0 0 0 1 1 1 0 0 0 0 0 0 0 1 1 1 0 1 1 0 \\
1 1 0 0 1 1 0 1 1 0 1 1 1 0 0 0 0 0 0 0 1 1 1 0 0 0 1 1 1 0 0 0 0 0 0 0 1 1 1 0 1 1 \\
0 1 1 0 0 1 1 0 1 1 0 1 1 1 0 0 0 0 0 0 0 1 1 1 0 0 0 1 1 1 0 0 0 0 0 0 0 1 1 1 0 1 \\
1 0 1 1 0 1 1 1 0 1 1 0 1 1 1 1 0 0 0 0 0 0 1 1 1 0 0 0 1 1 1 0 0 0 0 0 0 0 1 1 1 0 \\
1 1 0 1 1 1 1 1 1 0 1 1 0 1 1 1 0 0 0 0 0 0 0 1 1 1 0 0 0 1 1 1 0 0 0 0 0 0 0 1 1 1 \\
0 1 1 0 1 1 1 0 1 1 0 1 1 0 1 1 1 0 0 0 0 0 0 0 1 1 1 0 0 0 1 1 1 0 0 0 0 0 0 0 1 1 \\
1 0 1 1 0 1 1 0 0 1 1 0 1 1 0 1 1 1 0 0 0 0 0 0 0 1 1 1 0 0 0 1 1 1 0 0 0 0 0 0 0 1 \\
1 1 0 1 1 0 1 1 0 0 1 1 0 1 1 0 1 1 1 0 0 0 0 0 0 0 1 1 1 0 0 0 1 1 1 0 0 0 0 0 0 0 \\
1 1 1 0 1 1 0 1 1 1 0 1 1 0 1 1 0 1 1 1 0 0 0 0 0 0 0 1 1 1 0 0 0 1 1 1 0 0 0 0 0 0 \\
0 1 1 1 0 1 1 0 1 0 1 0 1 1 0 1 1 0 1 1 1 0 0 0 0 0 0 0 1 1 1 0 0 0 1 1 1 0 0 0 0 0 \\
0 0 1 1 1 0 1 1 0 1 1 0 0 1 1 0 1 1 0 1 1 1 0 0 0 0 0 0 0 1 1 1 0 0 0 1 1 1 0 0 0 0 \\
0 0 0 1 1 1 0 1 1 0 1 1 0 0 1 1 0 1 1 0 1 1 1 0 0 0 0 0 0 0 1 1 1 0 0 0 1 1 1 0 0 0 \\
0 0 0 0 1 1 1 0 1 1 0 1 1 1 1 1 1 0 1 1 0 1 1 1 0 0 0 0 0 0 0 1 1 1 0 0 0 1 1 1 0 0 \\
0 0 0 0 0 1 1 1 0 1 1 0 1 1 1 1 1 1 0 1 1 0 1 1 1 0 0 0 0 0 0 0 1 1 1 0 0 0 1 1 1 0 \\
0 0 0 0 0 0 1 1 1 0 1 1 0 1 1 0 0 1 1 0 1 1 0 1 1 1 0 0 0 0 0 0 0 1 1 1 0 0 0 1 1 1 \\
0 0 0 1 0 0 0 1 1 1 0 1 1 0 1 1 0 0 1 1 0 1 1 0 1 1 1 0 0 0 0 0 0 0 1 1 1 0 0 0 1 1 \\
1 0 0 0 0 0 0 0 1 1 1 0 1 1 0 1 1 0 1 1 1 0 1 1 0 1 1 1 0 0 0 0 0 0 0 1 1 1 0 0 0 1 \\
1 1 0 0 0 0 0 0 0 1 1 1 0 1 1 0 1 1 1 1 1 1 0 1 1 0 1 1 1 0 0 0 0 0 0 0 1 1 1 0 0 0 \\
0 1 1 0 0 0 0 0 0 0 1 1 1 0 1 1 0 1 1 1 0 1 1 0 1 1 0 1 1 1 0 0 0 0 0 0 0 1 1 1 0 0 
\end{tabular} \end{singlespacing}
\end{table}

\begin{table}[h]
\caption{{\bf The coloring matrix shown in Fig. \ref{fig7}d with only 2 monochromatic 
5-cliques sitting on 6 vertices of a complete graph with 43 nodes ($E_{min} = 2$). }} 
\label{tableS16} \begin{singlespacing}  \sf
\begin{tabular}{l} 
0 1 1 1 0 0 0 0 1 1 1 0 1 0 1 0 0 1 0 0 0 1 1 0 0 0 1 0 0 1 0 1 0 1 1 1 0 0 0 0 1 1 1 \\
1 0 1 1 1 0 0 1 0 1 1 1 0 1 0 1 0 0 1 0 0 0 1 1 0 0 0 1 0 0 1 0 1 0 1 1 1 0 0 0 0 1 1 \\
1 1 0 1 1 1 0 0 1 0 1 1 1 0 1 0 1 0 0 1 0 0 0 1 1 0 0 0 1 0 0 1 0 1 0 1 1 1 0 1 0 0 1 \\
1 1 1 0 1 1 1 0 0 1 0 1 1 1 0 1 0 1 0 0 1 0 0 0 1 1 0 0 0 1 0 0 1 0 1 0 1 1 1 0 1 0 0 \\
0 1 1 1 0 1 1 1 0 0 0 0 1 1 1 0 1 0 1 0 0 1 0 0 0 1 1 0 0 0 1 0 0 1 0 1 0 1 1 1 0 1 0 \\
0 0 1 1 1 0 1 1 1 0 0 0 0 1 1 1 0 1 0 1 0 0 1 0 0 0 1 1 0 0 0 1 0 0 1 0 1 0 1 1 1 0 0 \\
0 0 0 1 1 1 0 1 1 1 0 0 0 0 1 1 1 0 1 0 1 0 0 1 0 0 0 1 1 0 0 0 1 0 0 1 0 1 0 1 1 1 0 \\
0 1 0 0 1 1 1 0 1 1 1 0 0 1 0 1 1 1 0 1 0 1 0 0 1 0 0 0 1 1 0 0 0 1 0 0 1 0 1 0 1 1 1 \\
1 0 1 0 0 1 1 1 0 1 1 1 0 0 1 0 1 1 1 0 1 0 1 0 0 1 0 0 0 1 1 0 0 0 1 0 0 1 0 1 0 1 1 \\
1 1 0 1 0 0 1 1 1 0 1 1 1 0 0 1 0 1 1 1 0 1 0 1 0 0 1 0 0 0 1 1 0 0 0 1 0 0 1 0 1 0 1 \\
1 1 1 0 0 0 0 1 1 1 0 1 1 1 0 0 0 0 1 1 1 0 1 0 1 0 0 1 0 0 0 1 1 0 0 0 1 0 0 1 0 1 0 \\
0 1 1 1 0 0 0 0 1 1 1 0 1 1 1 0 0 0 0 1 1 1 0 1 0 1 0 0 1 0 0 0 1 1 0 0 0 1 0 0 1 0 1 \\
1 0 1 1 1 0 0 0 0 1 1 1 0 1 1 1 0 0 1 0 1 1 1 0 1 0 1 0 0 1 0 0 0 1 1 0 0 0 1 0 0 1 0 \\
0 1 0 1 1 1 0 1 0 0 1 1 1 0 1 1 1 0 0 1 0 1 1 1 0 1 0 1 0 0 1 0 0 0 1 1 0 0 0 1 0 0 1 \\
1 0 1 0 1 1 1 0 1 0 0 1 1 1 0 1 1 1 0 0 1 0 1 1 1 0 1 0 1 0 0 1 0 0 0 1 1 0 0 0 1 0 0 \\
0 1 0 1 0 1 1 1 0 1 0 0 1 1 1 0 1 1 1 0 0 0 0 1 1 1 0 1 0 1 0 0 1 0 0 0 1 1 0 0 0 1 0 \\
0 0 1 0 1 0 1 1 1 0 0 0 0 1 1 1 0 1 1 1 0 0 0 0 1 1 1 0 1 0 1 0 0 1 0 0 0 1 1 0 0 0 1 \\
1 0 0 1 0 1 0 1 1 1 0 0 0 0 1 1 1 0 1 1 1 0 0 1 0 1 1 1 0 1 0 1 0 0 1 0 0 0 1 1 0 0 0 \\
0 1 0 0 1 0 1 0 1 1 1 0 1 0 0 1 1 1 0 1 1 1 0 0 1 0 1 1 1 0 1 0 1 0 0 1 0 0 0 1 1 0 0 \\
0 0 1 0 0 1 0 1 0 1 1 1 0 1 0 0 1 1 1 0 1 1 1 0 0 1 0 1 1 1 0 1 0 1 0 0 1 0 0 0 1 1 0 \\
0 0 0 1 0 0 1 0 1 0 1 1 1 0 1 0 0 1 1 1 0 1 1 1 0 0 0 0 1 1 1 0 1 0 1 0 0 1 0 0 0 1 1 \\
1 0 0 0 1 0 0 1 0 1 0 1 1 1 0 0 0 0 1 1 1 0 1 1 1 0 0 0 0 1 1 1 0 1 0 1 0 0 1 0 0 0 1 \\
1 1 0 0 0 1 0 0 1 0 1 0 1 1 1 0 0 0 0 1 1 1 0 1 1 1 0 0 0 0 1 1 1 0 1 0 1 0 0 1 0 0 0 \\
0 1 1 0 0 0 1 0 0 1 0 1 0 1 1 1 0 1 0 0 1 1 1 0 1 1 1 0 0 1 0 1 1 1 0 1 0 1 0 0 1 0 0 \\
0 0 1 1 0 0 0 1 0 0 1 0 1 0 1 1 1 0 1 0 0 1 1 1 0 1 1 1 0 0 1 0 1 1 1 0 1 0 1 0 0 1 0 \\
0 0 0 1 1 0 0 0 1 0 0 1 0 1 0 1 1 1 0 1 0 0 1 1 1 0 1 1 1 0 0 1 0 1 1 1 0 1 0 1 0 0 1 \\
1 0 0 0 1 1 0 0 0 1 0 0 1 0 1 0 1 1 1 0 0 0 0 1 1 1 0 1 1 1 0 0 0 0 1 1 1 0 1 0 1 0 0 \\
0 1 0 0 0 1 1 0 0 0 1 0 0 1 0 1 0 1 1 1 0 0 0 0 1 1 1 0 1 1 1 0 0 0 0 1 1 1 0 1 0 1 0 \\
0 0 1 0 0 0 1 1 0 0 0 1 0 0 1 0 1 0 1 1 1 0 0 0 0 1 1 1 0 1 1 1 0 0 1 0 1 1 1 0 1 0 1 \\
1 0 0 1 0 0 0 1 1 0 0 0 1 0 0 1 0 1 0 1 1 1 0 1 0 0 1 1 1 0 1 1 1 0 0 1 0 1 1 1 0 1 0 \\
0 1 0 0 1 0 0 0 1 1 0 0 0 1 0 0 1 0 1 0 1 1 1 0 1 0 0 1 1 1 0 1 1 1 0 0 1 0 1 1 1 0 1 \\
1 0 1 0 0 1 0 0 0 1 1 0 0 0 1 0 0 1 0 1 0 1 1 1 0 1 0 0 1 1 1 0 1 1 1 0 0 0 0 1 1 1 0 \\
0 1 0 1 0 0 1 0 0 0 1 1 0 0 0 1 0 0 1 0 1 0 1 1 1 0 0 0 0 1 1 1 0 1 1 1 0 0 0 0 1 1 1 \\
1 0 1 0 1 0 0 1 0 0 0 1 1 0 0 0 1 0 0 1 0 1 0 1 1 1 0 0 0 0 1 1 1 0 1 1 1 0 0 1 0 1 1 \\
1 1 0 1 0 1 0 0 1 0 0 0 1 1 0 0 0 1 0 0 1 0 1 0 1 1 1 0 1 0 0 1 1 1 0 1 1 1 0 0 1 0 1 \\
1 1 1 0 1 0 1 0 0 1 0 0 0 1 1 0 0 0 1 0 0 1 0 1 0 1 1 1 0 1 0 0 1 1 1 0 1 1 1 0 0 1 0 \\
0 1 1 1 0 1 0 1 0 0 1 0 0 0 1 1 0 0 0 1 0 0 1 0 1 0 1 1 1 0 1 0 0 1 1 1 0 1 1 1 0 0 0 \\
0 0 1 1 1 0 1 0 1 0 0 1 0 0 0 1 1 0 0 0 1 0 0 1 0 1 0 1 1 1 0 0 0 0 1 1 1 0 1 1 1 0 0 \\
0 0 0 1 1 1 0 1 0 1 0 0 1 0 0 0 1 1 0 0 0 1 0 0 1 0 1 0 1 1 1 0 0 0 0 1 1 1 0 1 1 1 0 \\
0 0 1 0 1 1 1 0 1 0 1 0 0 1 0 0 0 1 1 0 0 0 1 0 0 1 0 1 0 1 1 1 0 1 0 0 1 1 1 0 1 1 1 \\
1 0 0 1 0 1 1 1 0 1 0 1 0 0 1 0 0 0 1 1 0 0 0 1 0 0 1 0 1 0 1 1 1 0 1 0 0 1 1 1 0 1 1 \\
1 1 0 0 1 0 1 1 1 0 1 0 1 0 0 1 0 0 0 1 1 0 0 0 1 0 0 1 0 1 0 1 1 1 0 1 0 0 1 1 1 0 1 \\
1 1 1 0 0 0 0 1 1 1 0 1 0 1 0 0 1 0 0 0 1 1 0 0 0 1 0 0 1 0 1 0 1 1 1 0 0 0 0 1 1 1 0
\end{tabular} \end{singlespacing}
\end{table}

\vfill

\vfill

\end{document}